\documentclass[twocolumn]{aastex631}

\usepackage{amssymb}
\usepackage{amsmath}
\usepackage{amsfonts}
\usepackage{bm}
\usepackage{graphicx}
\usepackage{subfigure}
\usepackage{calc}
\usepackage{epsfig}
\usepackage{url}

\usepackage{hyperref}

\newcommand{\be}{\begin{equation}}
\newcommand{\ee}{\end{equation}}
\newcommand{\bs}{\begin{split}}
\newcommand{\es}{\end{split}}
\newcommand{\R}[1]{\textcolor{red}{ #1}}
\newcommand{\B}[1]{\textcolor{blue}{#1}}

\begin{document}

\title{The Role of Population III Star Tidal Disruption Events in Black Hole Growth at the Cosmic Dawn}

\author{Zijian Wang}
\affiliation{National Gravitation Laboratory, Hubei Key Laboratory of Gravitation and Quantum Physics, School of Physics, Huazhong University of Science and Technology, Luoyu Road 1037, Wuhan, China}
\author{Yiqiu Ma}
\altaffiliation{myqphy@hust.edu.cn}
\affiliation{National Gravitation Laboratory, Hubei Key Laboratory of Gravitation and Quantum Physics, School of Physics, Huazhong University of Science and Technology, Luoyu Road 1037, Wuhan, China}
\affiliation{Department of Astronomy, School of Physics, Huazhong University of Science and Technology, Luoyu Road 1037, Wuhan, China}
\author{Yuxuan Li}
\affiliation{National Gravitation Laboratory, Hubei Key Laboratory of Gravitation and Quantum Physics, School of Physics, Huazhong University of Science and Technology, Luoyu Road 1037, Wuhan, China}
\author{Zheng Cai}
\affiliation{Department of Astronomy, Tsinghua University, Beijing 100084, People's Republic of China}
\author{Chanyan Wang}
\affiliation{National Gravitation Laboratory, Hubei Key Laboratory of Gravitation and Quantum Physics, School of Physics, Huazhong University of Science and Technology, Luoyu Road 1037, Wuhan, China}
\author{Qingwen Wu}
\affiliation{Department of Astronomy, School of Physics, Huazhong University of Science and Technology,
Luoyu Road 1037, Wuhan, China}

\begin{abstract}
The discovery of supermassive black holes (SMBHs) at high redshifts has intensified efforts to understand their early formation and rapid growth during the cosmic dawn. Using a semi-analytical cosmological framework, we investigate the role of tidal disruption events (TDEs) involving Population III (Pop-III) stars in driving the growth of heavy seed black holes ($10^4-10^6\,M_{\odot}$). Our results indicate that Pop-III TDEs significantly accelerate the growth of relatively lighter massive black holes ($\sim 10^4-10^5\,M_{\odot}$), allowing them to increase their mass by roughly an order of magnitude within the first 10 Myr. Cosmological evolution modeling further supports that such Pop-III TDE-driven growth scenarios are consistent with the formation pathways of observed luminous high-redshift quasars originating from seed black holes at $10<z<15$. We also discuss the future observational probes of these early-stage growth processes that future facilities, including space-based gravitational wave observatories and infrared telescopes like JWST, could potentially detect. These findings provide a clear observational framework to test the critical role of Pop-III star interactions in the rapid buildup of SMBHs during the earliest epochs.
\end{abstract}


\section{Introduction} \label{sec:intro}

Supermassive Black Holes (SMBHs) reside at the centers of nearly all galaxies, as strongly supported by the observational demography of local galaxies. Understanding the evolution history of these SMBHs is one of the most important tasks of modern astrophysics. The detection of quasars at $z\sim 6$ suggests that these SMBHs may orginated from the massive seed black holes (BHs) at the Cosmic Dawn\,($z\sim 10-15$) \citep{volonteri2021}. Currently, the properties and the growth process of these seed BHs are still poorly understood and not observationally accessible yet.

Seed BHs can form through the gravitational collapse of Population-III (Pop-III) stars---the first generation of stars in the universe---in a process known as the light-seed channel \citep{madau_rees2001,schneider2002}. However, it isn't easy that these light seed BHs, with initial masses of a few tens to hundreds of solar masses, can grow to $10^9\,M_\odot$ within less than a gigayear \citep{pfister2019,ma2021}. To address this, an alternative formation channel for more massive seed BHs, with masses of 
$10^4-10^6\,M_\odot$, has been proposed \citep{bromm_loeb2003,BVR2006,LN2006,shang2010,montero2012,begelman2008}. These heavy seed BHs, often referred to as direct-collapse black holes (DCBHs), form through the rapid collapse of pre-galactic gas in atomic cooling haloes, which is supported by the Lyman-Werner photon feedback \citep{haiman1997,Dijkstra2008,Dijkstra2014}. Besides, heavy seed BHs may also form via super-Eddington accretion onto light seed BHs \citep{lupi2014,lupi2016,shi2023,mehta2024}, though this scenario is not the focus of our discussion.

The key physical process connecting seed black holes (BHs) to their later massive counterparts is the growth mechanism. The growth of seed black holes (BHs) can occur through the accretion of surrounding gas as well as binary black hole (BBH) mergers. Futhermore, Tidal Disruption Events (TDEs) provide an additional channel for black hole growth \citep{hills1975}. After the formation of the DCBH, the surrounding nuclear accretion disk is gravitationally unstable, hence causing the metal-poor gas to fragment into clumps that later evolve into Population III (Pop III) stars. Theoretical studies suggest these stars were typically an order of magnitude more massive than their Population I/II (Pop-I/II) counterparts \citep{Hirano2014,Hirano2015}. Through TDEs, Pop III stars can transfer their stellar mass to the central black hole, with each event potentially supplying the nascent SMBH seed with $\sim10^2\,M_\odot$ of mass. Previous studies have explored the initial growth of intermediate-mass black holes (IMBHs) through Population I/II TDEs, suggesting that this growth mechanism is efficient during their early evolutionary stages \citep{milosavljevic2006,alexander2017,Rizzuto2023,pfister2021,lee2023}. In this work, we want to explore how the Tidal Disruption Events involving Pop-III stars and DCBH can affect the seed BH growth, despite the modelling uncertainty due to the complicated nature of the scenarios. Our approach further combines the modeling of Pop-III star TDE around DCBHs with semi-analytical cosmological simulations based on the Extended Press-Schechter formalism. We find that Pop-III TDEs boost DCBH's growth in the first 10 Myr. This enhances the formation of the high-redshift AGNs and modifies the evolution of DCBH mass function.

Despite the challenges in probing the properties of seed BHs, observational astronomy is making significant strides into the high-redshift universe. Cutting-edge facilities like the James Webb Space Telescope (JWST), equipped with state-of-the-art infrared instruments (e.g., NIRCam, NIRSpec and MIRI), are set to revolutionize our view of the early universe, offering potential capabilities that are suited to probing the elusive properties of seed black holes \citep{pacucci2015,natarajan2017,barrow2018,Valiante2018,whalen2020,trinca2023}. Additionally, future gravitational wave (GW) detectors, both ground-based and spaceborne, will offer powerful tools to study these BHs. Third-generation ground-based detectors like the Einstein Telescope and Cosmic Explorer will be capable of detecting $30\,M_\odot+30\,M_\odot$ binary black hole (BBH) systems up to redshifts of $z\sim 100$ \citep{ET,CE}. Meanwhile, spaceborne detectors such as LISA/Tianqin/Taiji will precisely measure the masses of coalescing BBH systems in the range of $10^4-10^7\,M_\odot$ up to $z\sim 20$ \citep{LISA,tianqin,taiji}. These observations will play a pivotal role in unraveling the physical processes associated with the formation and evolution of seed BHs. In this work, the possible observational probes to the growth of the seed black hole boosted by Pop-III TDE will also be discussed.

The structure of this paper is organized as follows. Section\,\ref{sec:general} introduces the general physical scenario of our work and Section\,\ref{sec:methods} presents the two key methods that are implemented in our analysis: how the Pop-III TDE around DCBH is modelled, and how we combine this model with the cosmological simulation. Then Section\,\ref{sec:results} is devoted to analyzing the results of our work and finally we conclude the paper in Section\,\ref{sec:conclusion}.

\section{General physical picture} \label{sec:general}
Before presenting the details of our analysis, we first introduce the general physical picture of our system. 

DCBHs form in atomic-cooling haloes, where intense Lyman-Werner radiation from nearby star-forming galaxies or haloes dissociates molecular hydrogen (H$_2$) in the pristine gas. This process suppresses efficient H$_2$-cooling, allowing less-efficient atomic hydrogen cooling to dominate. As a result, more massive gas clouds can accumulate, facilitating the formation of DCBHs.  

\begin{figure}
    \centering
    \includegraphics[width=1.0\linewidth]{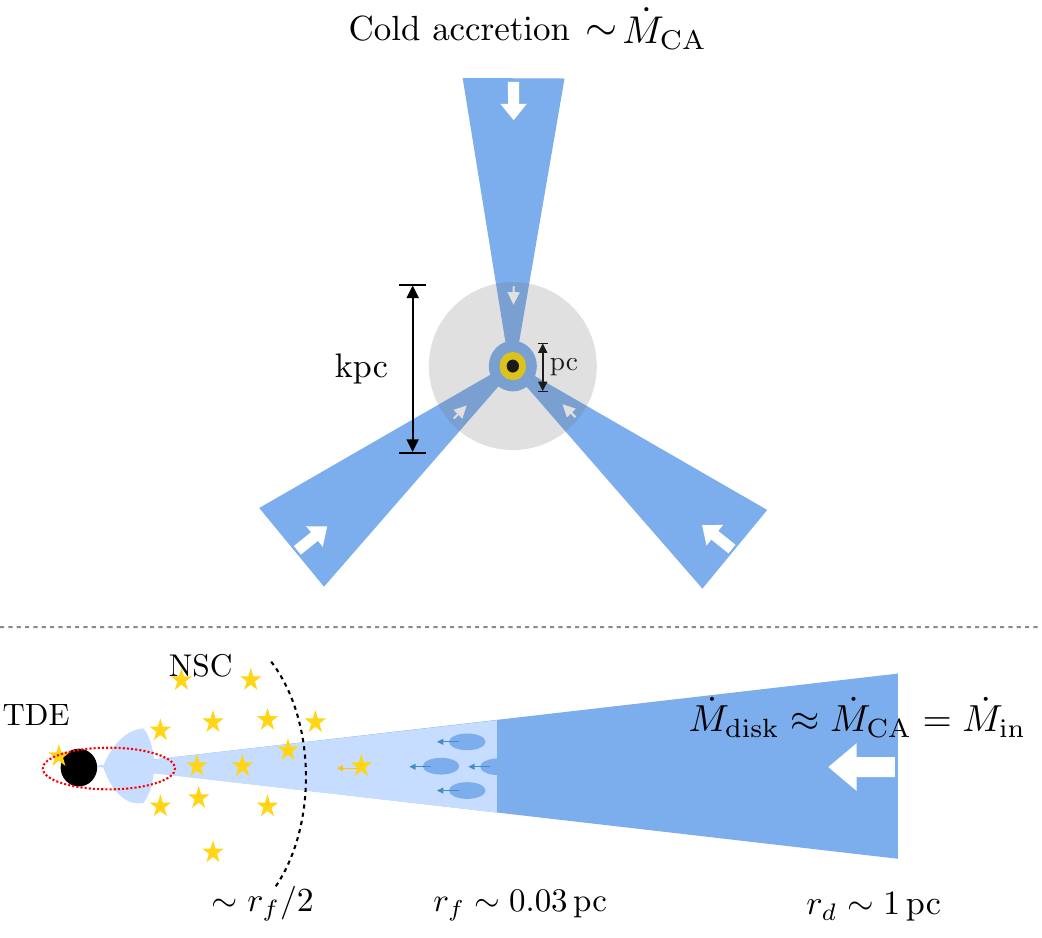}
    \caption{Schematic picture outlines the physical picture at the cosmic dawn. Upper panel: Cold accretion gas from the cosmic web flows into the halo center via the filament structure, where the grey, blue, yellow, and black circles represent the dark matter halo, central disk, fragmentation region, and the central DCBH, respectively. Lower panel: Physical process happen in the central disk, including gas fragmentation into clumps, migration of clumps and stars, the formation of a nuclear star cluster, and the TDE. The accretion flow in the vicinity of the central BH forms a slim disk.}
        \label{fig:scheme}
    \end{figure}

The formed DCBH interacts with its surrounding environment, forming a structure schematically illustrated in Figure\,\ref{fig:scheme}, as first discussed in \citet{kashiyama2016}.  For a DCBH with mass $M_\mathrm{BH} \sim 10^5\,M_{\odot}$, accretion generates a nuclear accretion disk extending to approximately $r_{\rm edge}\sim 10^{-4}$\,pc. Beyond this disk lies a star-forming region, which can extend to the fragmentation radius $r_\mathrm{f}\sim 0.01-0.1$\,pc. Within this region, the gas fragments into clumps, collapses, and eventually forms metal-poor, massive Population III stars with masses in the range $m_*\sim 10-100\,M_{\odot}$. These Pop III stars can gravitationally interact, forming a nuclear star cluster, and some may scatter into the loss cone, leading to Pop-III TDEs. Outside the fragmentation radius, the gas forms a self-gravitating disk, where cooling is insufficient to trigger further fragmentation, allowing the gas to inflow steadily. 

The above picture illustrates the basic physical scenarios around one DCBH, while many haloes can host DCBHs and their corresponding Pop III TDEs at a specific redshift. Therefore a cosmological statistical analysis is necessary. Moreover, the generated DCBHs and their host halo will evolve with decreasing redshift via various growing mechanisms, e.g. gas/dark matter accretion, the TDEs discussed in this work, and binary merge, forming more massive BHs. As we shall elaborate in the next section, we generate the halo histories using the semi-analytical approach: Monte-Carlo merger trees based on the Extended Press-Schechter theory, following the algorithm presented in \citet{parkinson2008}. For haloes exposed to sufficiently intense Lyman-Werner radiation, the suppression of H$_2$-cooling allows them to be identified as potential hosts for DCBHs before H$_2$-cooling can become efficient.
We will trace the evolution of the DCBHs in these identified haloes, and examine how the Pop-III TDE process affect the growth process and the DCBH mass distribution.

\begin{table}[h]
\centering
\begin{tabular}{ccc}
\hline
\textbf{Symbol} & \textbf{Value} & \textbf{Description} \\
\hline
$M_\mathrm{BH}$ & $10^4-10^6~\mathrm{M_\odot}$ & Initial mass of DCBH \\
$\dot{M}_\mathrm{in}$ & $0.13~\mathrm{M_\odot}~\mathrm{yr}^{-1}$ & Rate of gas inflow \\
$r_\mathrm{f}$ & $0.02-0.03~\mathrm{pc}$ & Fragmentation radius\\
$\alpha$ & 0.17 & Pop-III IMF Power-index\\
$\langle m_*\rangle$ & $70~\mathrm{M_\odot}$ & Averaged stellar mass\\
$\gamma$ & $7/4$ & Stellar density cusp index \\
$\varepsilon$ & 0.1 & Star formation efficiency \\
$t_\mathrm{life}$ & $10~\mathrm{Myr}$ & Pop-III Star Lifetime\\
\hline
\end{tabular}
\caption{Parameters of the fiducial model.}
\label{tab:physics_parameters}
\end{table}

\section{Methods} \label{sec:methods}

In this section, we will describe the methods adopted for simulations, including the modelling of the Pop-III TDE around the DCBH, the approaches used to generate Monte Carlo dark matter halo merger trees, the criteria for DCBH formation, and the black hole mass growth process. We assume a $\Lambda$ Cold Dark Matter ($\Lambda$CDM) cosmology with parameters $\Omega_m=0.315,~\Omega_\Lambda=0.685,~\Omega_b=0.0493,~h=0.674.~\sigma_8=0.811$ \citep{planck2020}.

\subsection{Modelling Pop-{\rm III} TDEs}\label{sec:tde_modelling}

Tidal disruption events occur when a main-sequence star approaches a central black hole so closely that the BH's tidal forces overcome the star's self-gravity, leading to its disruption. For TDEs involving Pop-III stars and a DCBH, three key conditions must be met: (1) A sufficient number of Pop-III stars must exist in the vicinity of the DCBH to provide potential candidates for disruption. (2) These Pop-III stars must form a nuclear star cluster (NSC), creating an environment where gravitational scattering can drive stars into eccentric orbits conducive to TDEs. (3) These processes must occur within the lifetime of a Pop-III star; otherwise, supernova explosions will chemically contaminate the surrounding environment, altering its pristine conditions \citep{wise2012}. Note that we ignore the potential chemical contamination from TDE debris, as we argue that such debris will only affect the DCBH's immediate local environment.

\subsubsection{Baryonic gas supply}
In the early universe, supermassive stars (SMS) or DCBHs form in rare, massive dark matter haloes located at the intersections of cosmic filaments, where multiple high-density streams of gas and dark matter converge. In these regions, the gas and dark matter densities are as high as $\sim 10-10^2$ cosmic mean density. Due to the halo’s deep gravitational potential and suppression of cooling by strong Lyman-Werner radiation, these filamentary inflows channel pristine gas at rates of $\dot{M}_{\rm CA}$\,(typically $0.1-1\,M_{\odot}/{\rm yr}$), far exceeding typical accretion, to the halo center. This process is the so-called ``cold accretion", which typically emerges for halo with mass $10^{7-8}\,M_{\odot}$ \citep{DiMatteo2012}. Simulations show that gas from the cosmic web efficiently spirals inward, with gravitational torques and shocks redirecting angular momentum outward, enabling $90\%$ of the infalling gas to reach the central accretion disk with sub-parsec scale \citep{masaki2024}. The central accretion disk can self-regulate so that the gas supply rate from cold accretion $\dot{M}_{\rm CA}$ can always be balanced with the central disk's accretion rate $\dot{M}_{\rm disk}$. Hence we can write
\begin{equation}
\dot{M}_{\rm CA}\approx\dot{M}_{\rm disk}\equiv\dot{M}_{\mathrm{in}}=0.13\,M_\odot/{\rm yr} \left(\frac{T}{10^4\,\mathrm{K}}\right)^{3/2}.
\end{equation}
This hierarchical process, spanning kiloparsec-scale filament flows to sub-parsec disk feeding, creates the hyper-efficient fuel reservoir needed for SMS/DCBH formation. Over the typical lifetime of a Pop-III star $\sim 3-10$\,Myr \citep{schaerer2002}, the accumulated mass of the inflowing gas can reach $\sim 3\times 10^5-10^6\,M_\odot$. This mass is sufficient to form $\sim 10^3-10^4$ Pop-III stars, assuming a characteristic mass range of $10-100\,M_\odot$ per star. \\

\subsubsection{Nuclear star cluster}
As discussed earlier, cooling in the central accretion disk gas triggers Pop-III star formation. High-resolution simulations suggest three possible Pop-III initial mass function\,(IMF) slopes \,\citep{stacy2013, grief2011},
\be
P(M)\propto M^{-\alpha},\quad \alpha = -0.17,~0,~0.17,
\ee 
which distinct from the typical Salpeter IMF $\alpha = 2.35$\,\citep{salpeter1955}. We adopt two models: Model-1 with an averaged stellar mass $\langle m_*\rangle = 70\,\mathrm{M_\odot}$ \citep{Jaacks2018} and Model-2 with $\langle m_*\rangle = 150\,\mathrm{M_\odot}$ \citep{Chowdhury2024}. The stellar lifetimes in these models are typically 10 Myr and 3 Myr. On timescales longer than these lifetimes, Pop-III SNe produce a metallicity floor of $Z\sim 10^{-3} Z_\odot$ \citep{wise2012}, shifting subsequent star formation to the Pop-I/II regime.

Several time scales are important for the formation of the nuclear star cluster\,(NSC). The first one $t_*=\langle m_* \rangle/\epsilon \dot{M}_{\rm in} $ characterizes the time scale of the star formation gas supply. For Model-1 with $\langle m_* \rangle=70\,M_{\odot}$ and $\epsilon=0.1, \dot M_{\rm in}=0.13\,M_{\odot}/{\rm yr}$, we have $t_*\approx 5\times 10^3$\,yr, while  for more massive Model-2 we have $t_*\approx 1.1\times 10^4$\,yr.

At the fragmentation radius $r_\mathrm{f}$, represented by
\begin{equation}
r_\mathrm{f}=4.2\times10^{-4}\mathrm{pc}\left(\dfrac{M_\mathrm{BH}}{M_\odot}\right)^{1/3}\left(\dfrac{\dot{M}_\mathrm{in}}{\mathrm{M_\odot~yr^{-1}}}\right)^{-2/9},
\end{equation} 
($r_\mathrm{f} = 0.033\,\mathrm{pc}$ for a central black hole mass of $10^5 M_\odot$, \citet{kashiyama2016}), the inflowing gas starts to fragment into gas clumps. These gas clumps will migrate inward to the region of NSC ($\sim r_\mathrm{f}/2$), accrete more mass, and form stars. Two important time scales are relevant for this process: (1) the migration timescale of the gas clumps from $r_\mathrm{f}$ to $r_\mathrm{f}/2$ can be estimated as $t_{\rm mig}\sim 1.3\times 10^4$\,yrs for $10^5\,M_\odot$ central BH, and $5.4\times 10^3$\,yrs for slightly lighter $3\times 10^4\,M_\odot$ central BH.  (2) The star-forming Kelvin-Hemholtz timescale $t_{\rm KH}\sim \mathrm{several}~10^3\,{\rm yrs}\leq t_{\rm mig}$ \citep{OP2003}, which means that the gas clump will transit to Pop-III stars during its migration.

The inward-migrated stars will gravitationally mutual-interact and form the NSC. Specifically,   multi-gravitational interactions happen among the stars that are initially distributed in the disk plane. These interactions drive the relaxation of the stars' orbital eccentricity distribution and the inclination angles relative to the disk plane, with related timescales $\tau_e$ and $\tau_\theta$, respectively. These time scales are crucial for establishing the NSC's spherical geometry
and can be estimated as \citep{stewart2000,kocsis2011}:
\be
\tau_{\theta}\sim 2\tau_e\approx2\times 0.22\frac{\pi\langle e^2\rangle^2 (M_{\rm BH}r)^{3/2}}{G^{1/2}\langle m_*\rangle^2{\rm ln}\Lambda}\frac{1}{N_*}.
\ee
Here, $\langle e^2\rangle\sim0.09$ is the mean squared orbital eccentricity, $M_{\rm BH}$ is the mass of the central black hole, $r\sim r_\mathrm{f}/2$ is the NSC's characteristic radius, $\langle m_*\rangle$ is the average stellar mass, ${\rm ln}\Lambda$ is the Coulomb logarithm which accounts for the cumulative encounter effects in the gravitational scattering, and $N_*$ is the number of stars in the cluster. It is important to note that as the number of stars increases, these timescales become smaller. For $M_{\rm BH}\sim 10^5\,M_\odot$, $\langle m_*\rangle\approx 70\,M_\odot$, the NSC formation timescale $\tau_{\theta}$ takes the value $\sim [2\times 10^2, 4\times 10^3]\,{\rm yrs}$ 
when the star population $N_{*}\in[10,180]$. While for slightly lighter BH with $M_{\rm BH}\sim 3\times 10^4\,M_{\odot}$, we have $\tau_\theta\in[50, 1.5\times 10^3]\,{\rm yrs}$. This estimation shows that the disk relaxation time $\tau_\theta$ is much shorter than $t_{\rm mig}, t_*$.

Finally, the NSC formation timescale $t_{\rm cluster}\sim{\rm max}[t_{\rm mig}, t_*]\ll t_{\rm life}$, indicating that the NSC forms way before the supernova explosion of the Pop-III stars.

Furthermore, the change in the number of stars is a dynamical process influenced by star formation and consumption through TDEs, which can be described as:
\be
\frac{dN_*}{dt}=-\Gamma_{\rm TDE}(N_*)+\frac{1}{t_{\rm cluster}(N_*)},
\ee
where the $\Gamma_{\rm TDE}$ will be derived in the next subsection. Both TDE rate and the cluster star supply rate depends on the NSC star population $N_*$, therefore we can obtain the dynamical equilibrium point by numerically solving $dN_*/dt=0$. The resultant equlibrium NSC star population is $N_*\approx 140$ for $3\times 10^4\,M_\odot$ centeral BH and $N_*\approx 220$  for $10^5\,M_\odot$ central BH. Moreover, the radial number density distribution of these NSC stars is parametrized as
 \be
 n_*(r)=n_{*0}\left(\frac{r}{r_f}\right)^{-\gamma},
\ee
where the $\gamma$ is the cusp index. When the heavy stars are relatively common, $\gamma$ tends to be $7/4$, whereas in populations dominated by light stars, $\gamma$ typically falls within the range $3/2 < \gamma < 7/4$ \citep{alexander2009}.

Massive NSC stars can ionize the gas within the disk, and trigger photoevaporation at the disk surface via extreme ultraviolet \,(EUV) radiation. However, the size of the ionized region is much smaller than the disk radius hence, the outer disk remains neutral \citep{inayoshi2014}. Photoevaporation can in-principle prevent gas inflow and further affect the number of stars in the cluster \citep{kashiyama2016}.  The total evaporation rate of the disk is given by \citep{tanaka2013}:
\begin{equation}\label{eq:photoevaporation}
    \dot{M}_{\mathrm{pe}}=5.4\times 10^{-5}\left(\dfrac{\Phi_{\mathrm{EUV}}}{10^{49}\mathrm{s^{-1}}}\right)^{\frac{1}{2}}\left(\dfrac{r_\mathrm{d}}{1000\mathrm{AU}}\right)^{\frac{1}{2}}\mathrm{M_\odot~yr^{-1}}.
\end{equation}
In this formula, the $r_\mathrm{d}\sim 1~\mathrm{pc}$ is the disk radius\,\citep{regan2014} and the $\Phi_{\mathrm{EUV}}=N_*\Phi^*_{\mathrm{EUV}}$ represents EUV emissivity from all the newborn zero-age main sequence (ZAMS) stars, with $N_*$ the number of ZAMS and $\Phi^*=1.26\times10^{47}(M_*/\mathrm{M_\odot})^{1.4}~\mathrm{s^{-1}}$  the EUV emissivity for a single ZAMS star. Using the previously solved equilibrium NSC star population $N_*\approx 140-220$, the total EUV emissivity is given by $\Phi_{\mathrm{EUV}}\approx (0.67\sim 1)\times10^{52}~\mathrm{s^{-1}}$. Substituting this into Eq\,\eqref{eq:photoevaporation}, we obtain $\dot{M}_\mathrm{pe}\approx(2\sim 2.4)\times10^{-2}~\mathrm{M_\odot~yr^{-1}}$ and we have the ratio $\dot{M}_\mathrm{pe}/\dot{M}_{\mathrm{in}}\approx0.15-0.18\ll 1$, which means the EUV radiation is insufficient to deplete the gas supply onto the disk\,\footnote{It could be possible that the gas inflow can proceed even if $\dot{M}_\mathrm{pe}/\dot{M}_{\mathrm{in}}$ slightly exceeds the threshold \citep{inayoshi2014}}.


\subsubsection{Loss cone and tidal disruption event}
The gravitational interaction between the field stars in the NSC and an individual Pop-III star can scatter the star into the loss cone, leading to its eventual tidal disruption by the central DCBH. The standard loss-cone theory can describe this process, where the gravitational scattering dynamics is radius-dependent. Specifically, there exists a critical radius $r_{\rm crit}$ that separates the gravitational-scattering dynamics into the empty loss cone and full loss cone regions.  Its value is determined by the balance between the relaxation timescale $t_{\rm relax}$ and the dynamical timescale $t_{\rm dyn}$, which can be estimated as:
\be
r_{\rm crit}\sim \left[\frac{t_{\rm relax}(r)}{t_{\rm dyn}(r)}\right]^{1/(3+2\gamma)},
\ee
where the $t_{\rm dyn}\approx\sqrt{r^3/GM_{\rm BH}}\propto r^{3/2}$ represents the Keplerian dynamical timescale of the star.  The relaxation timescale by gravitational scattering $t_{\rm relax}$ is given by\,\citep{ch1943,spitzer1987}:
\be
\begin{split}
t_{\rm relax}&\approx0.34\frac{\sigma^3(r)}{G^2\langle m_*\rangle n_*(r){\rm ln}\Lambda}\\
&=\frac{0.34(GM_{\rm BH})^{3/2}r_f^\gamma}{G^2n_{*0}\langle m_*\rangle{\rm ln}\Lambda}r^{-3/2-\gamma},
\end{split}
\ee
with $\sigma(r)$ the velocity dispersion\,(typically equal to the Keplerian velocity). The full/empty loss cone region is determined by the ratio $t_{\rm dyn}/t_{\rm relax}\sim r^{3+\gamma}$.

In the full loss cone region (at large radii when the index $\gamma>-3$), the relaxation timescale is much shorter than the dynamical timescale. As a result, stars can efficiently scatter into the loss cone after just a few gravitational encounters. This leads to a high rate of TDEs because the loss cone is continuously repopulated.
In the empty loss cone region (at small radii), the relaxation timescale is much longer than the dynamical timescale. Stars require a large number of scatterings to enter the loss cone, and the supply of stars to the loss cone is limited. This results in a low rate of TDEs.   In our DCBH system with $M_\mathrm{BH}\sim10^5~\mathrm{M_\odot}$, the typical critical radius is $0.03~\mathrm{pc}$ (for $M_\mathrm{BH}\sim3\times10^4~\mathrm{M_\odot}$, we have $r_\mathrm{crit}\sim0.017~\mathrm{pc}$), which is comparable to the $r_\mathrm{f}$. Therefore, to have a correct estimation of the TDE rate, we need to take into account  its radii-dependence \citep{Merritt2013}:
\be
\begin{split}
&\Gamma_{\rm TDE}(r)=\\
&\int^{r_\mathrm{f}}_{r_t}dr\left[\frac{d\Gamma_{\rm empty}}{dr}f(q(r))+\frac{d\Gamma_{\rm full}}{dr}\left[1-f(q(r)\right]\right],
\end{split}
\ee
where
\be
\begin{split}
&\frac{d\Gamma_{\rm empty}}{dr}=\frac{4\pi r^2\rho(r)}{t_{\rm relax}(r)}{\rm ln}^{-1}\left[\frac{J_c^2}{J^2_{\rm LC}}\right],\\
&\frac{d\Gamma_{\rm full}}{dr}=\frac{4\pi r^2\rho(r)}{t_{\rm dyn}(r)}\left(\frac{r_t}{r}\right),
\end{split}
\ee
and the tidal disruptipn radius $r_t = R_{*} \left(M_{\rm BH}/m_{*}\right)^{1/3}$. The loss-cone angular momentum and the orbital angular momentum are $J_{\text{lc}} = \sqrt{2 G M_{\rm BH} r_t}$ and $J_c = \sqrt{G M_{\mathrm{BH}} r}$, respectively. The transition parameter $q(r)$ is defined as $q(r) =[ t_{\text{dyn}}(r)/t_{\text{relax}}(r) ]\left(J_c/J_{\rm lc}\right)^2$ and $f(q(r)) = 1/(1+q(r))$. 

\begin{figure}
    \centering
    \includegraphics[width=1.0\linewidth]{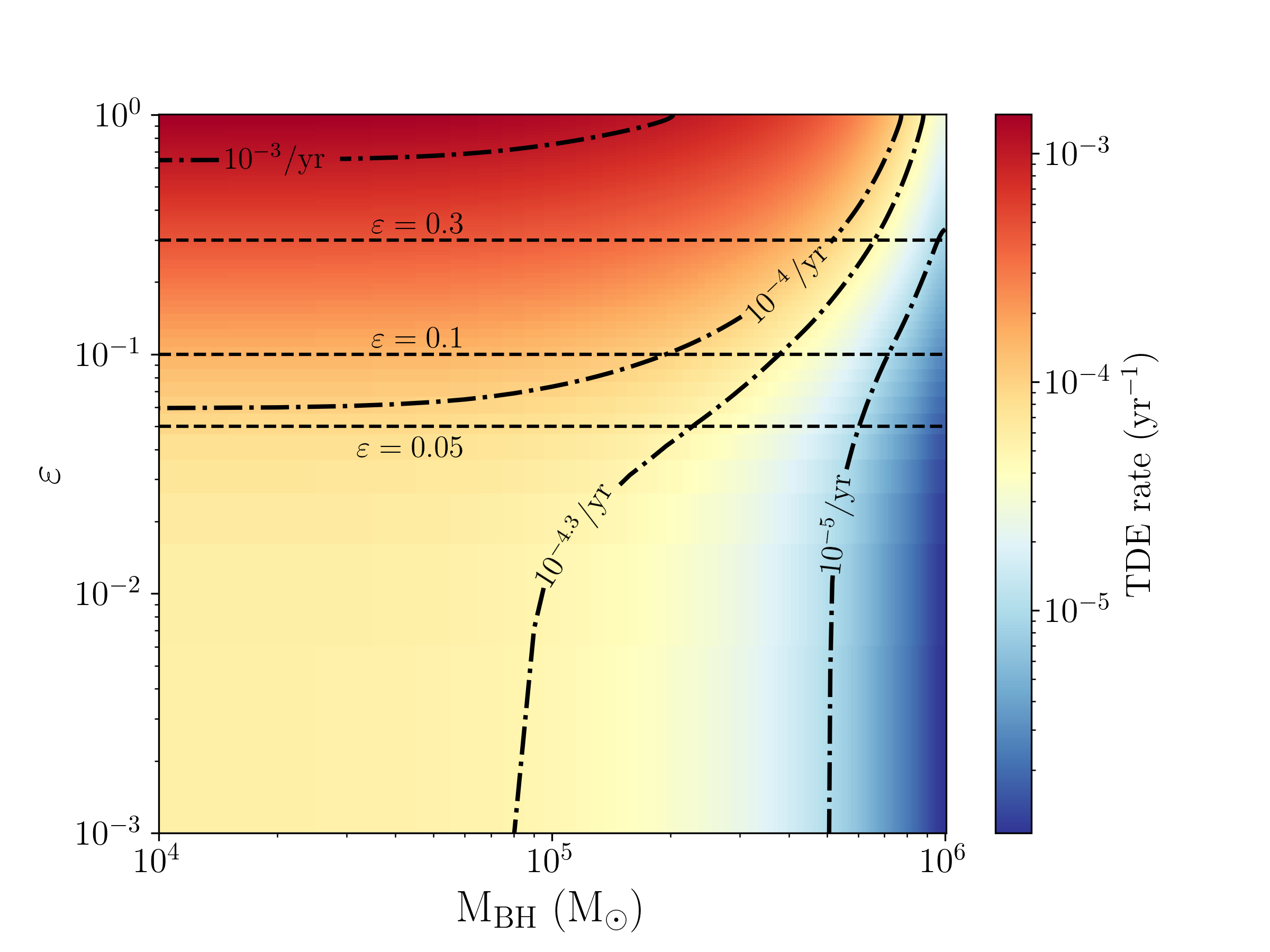}
    \caption{TDE rates for different central black hole masses (in the units of solar mass) and star formation efficiency $\varepsilon$, assuming a fiducial model with $\gamma = 7/4$ and Pop III IMF with a slope of $\alpha=0.17$.}
    \label{fig:m_eff_TDErate}
\end{figure}

\begin{figure}
    \centering
    \includegraphics[width=1.0\linewidth]{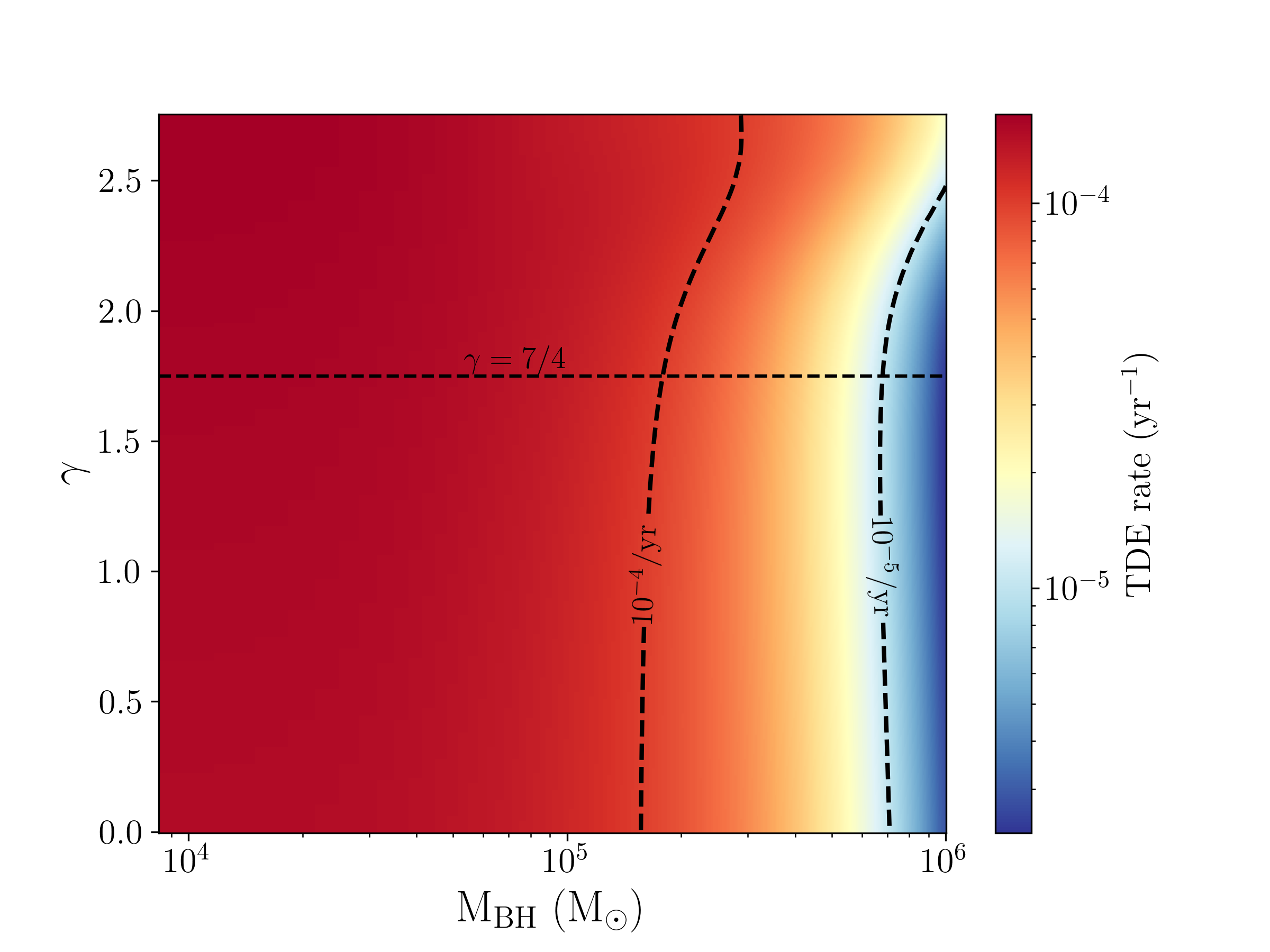}
    \caption{TDE rates for different central black hole masses (in the units of solar mass) and cusp index $\gamma$, assuming a fiducial model with $\varepsilon = 0.1,~\alpha = 0.17$.}
    \label{fig:m_gamma_TDErate}
\end{figure}

\begin{figure}
    \centering
    \includegraphics[width=1.0\linewidth]{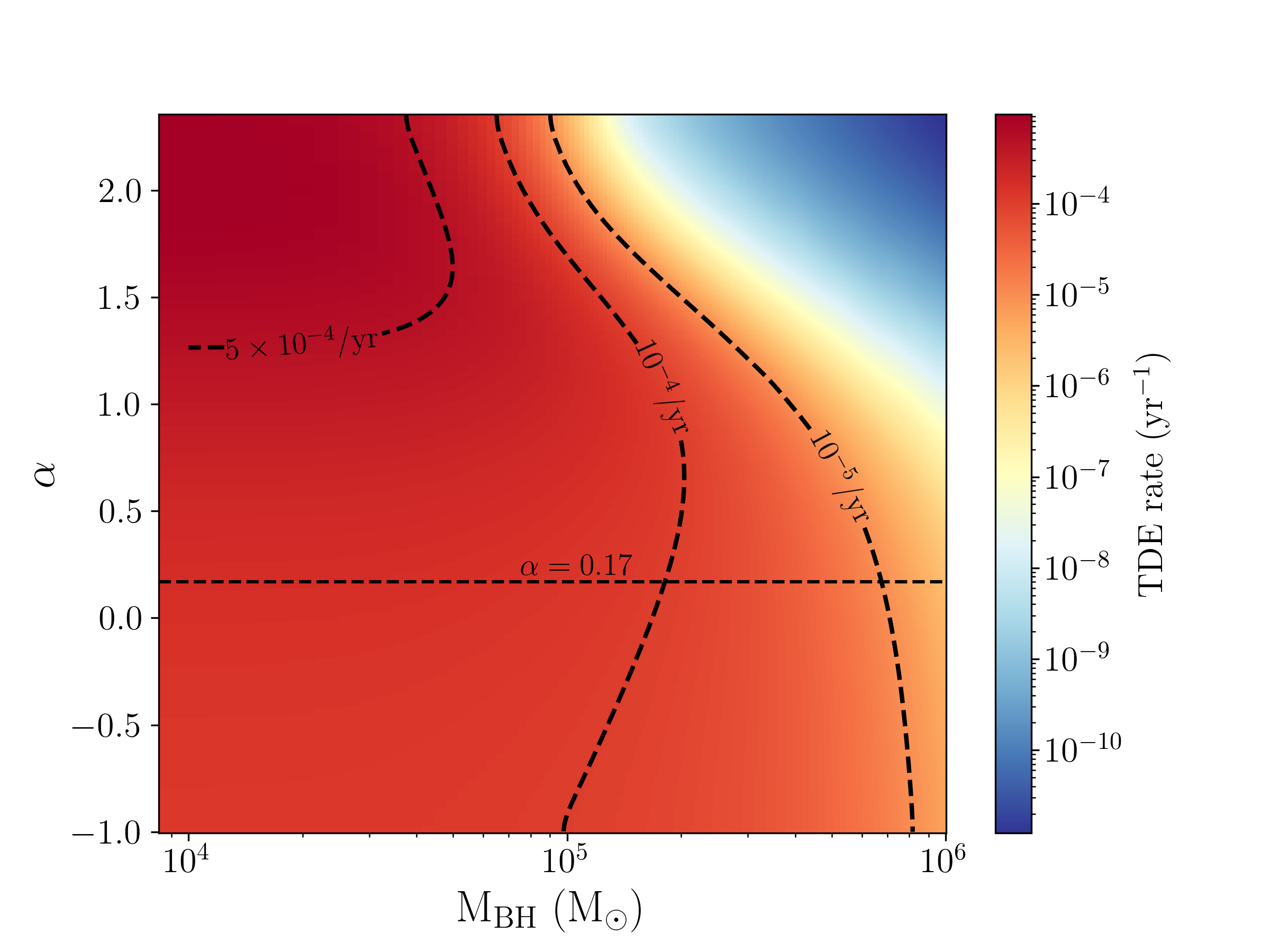}

    \caption{TDE rates for different central black hole masses (in the units of solar mass) and the power-law index $\alpha$ of Pop III initial mass function $P(M)\propto M^{-\alpha}$, assuming a fiducial model with $\varepsilon = 0.1,~\gamma=7/4$.}
    \label{fig:m_alpha_TDErate}
\end{figure}

In summary, the TDE rate in the vicinity of the DCBH is determined mainly by the following model-dependent quantities: the mass of the DCBH, the star formation efficiency ($\varepsilon$, SFE), the IMF of the Pop-III star\,($\alpha$-index), and the star's number density profile\,($\gamma$-index). The uncertainty of these quantities will affect the contribution of Pop-III TDE to the DCBH's growth, as demonstrated in Figures\,\ref{fig:m_eff_TDErate}, \ref{fig:m_gamma_TDErate} and \ref{fig:m_alpha_TDErate}.  Figure\,\ref{fig:m_gamma_TDErate} show that the cusp index $\gamma$ has only a minor effect on TDE rates. Similarly, as demonstrated in Figure\,\ref{fig:m_alpha_TDErate}, the power-law index $\alpha$ of Pop-III IMF lower than 0.17 also has a negligible impact. Although a higher $\alpha$ value would yield a larger TDE rate, the corresponding decrease in average stellar masses results in a comparable mass contribution to BH growth as the $\alpha=0.17$ case. In contrast,  a lower SFE leads to a significantly reduced TDE rates. However, recent JWST observations favor high SFE values in the high-redshift Universe \citep{shen2023,Andalman2024}. Therefore, we adopt $\gamma=7/4,\,\alpha=0.17,\,\varepsilon=0.1$ as the fiducial parameters in our subsequent simulations.


\subsection{Cosmological evolution of the Massive Black Holes}


Tracing the formation and growth of DCBHs through merger trees is a crucial step in understanding the cosmological evolution of massive black holes (MBHs). In this work, we utilize a semi-analytical model grounded in the extended Press-Schechter formalism \citep{PS, bond1991, bower1991, LC1993} to generate Monte Carlo realizations of dark matter halo merger trees. Additionally, we implement the algorithm outlined in \citet{parkinson2008}, which is an adapted version of the code used for dark matter halo history reconstructions in the \texttt{GALFORM} framework \citep{cole2000}. We generate $10^3$ dark matter halo merger trees extending down to 
redshift $z_{\rm max}=35$ with an adaptively high time resolution. The parent halo is assigned a mass of $10^{12}\,M_{\odot}$ at $z=6$, consistent with observational data of high-redshift quasars \citep{wyithe2006}.

\subsubsection{Identifying DCBHs in a Merger Tree}
The method we used for identifying DCBHs in the merger tree of dark matter haloes generally follows the approach developed by \citet{scoggins2024}, which is reviewed here for a complete illustration of the method used in this work. The first step involves identifying the host haloes. Since DCBHs typically form through the direct collapse of supermassive stars with intermediate masses around $\sim 10^5\,M_\odot$, and these supermassive stars can only form when the virial temperature of the halo reaches the atomic cooling threshold $T_{\rm vir}>10^4$\,K, we begin by selecting all dark matter haloes satisfying this condition, using the virial temperature formula \citep{barkana_loeb2001}:
\be\label{eq:virial_temperature}
\begin{split}
&T_{\text{vir}}\approx 1.98 \times 10^4 \, \text{K}\times \\
& \, \left[ \frac{\mu}{0.6} \right] \left[ \frac{M_{\text{vir}}}{10^{8} \, h^{-1} M_\odot} \right]^{2/3} \left[ \frac{\Omega_m}{\Omega_m(z)} \frac{\delta_c}{18\pi^2} \right]^{1/3} (\dfrac{1 + z}{10}),
\end{split}
\ee
where \( \mu \) is the mean molecular weight of the gas (typically \( \mu \approx 1.22 \) for  neutral primordial gas), \( M_{\text{vir}} \) is the virial mass of the halo, \( h \) is the Hubble parameter in units of \( 100 \, \text{km/s/Mpc} \),  \( \Omega_m(z) \) is the matter density parameter at redshift \( z \)\,(here $\Omega_m=\Omega_m(z=0)$), and \( \delta_c \) is the overdensity relative to the critical density (typically \( \delta_c \approx 200 \) for a virialized halo).

The next step involves tracing the progenitors of these selected haloes. For a halo with  $T_{\rm vir}\geq 10^4$\,K at a time slice corresponding to redshift $z_i$, we trace backward in time (toward increasing redshift) and examine its progenitor haloes at $z_i+\Delta z$. If any progenitor halo satisfies $T_{\rm vir}\geq 10^4$\,K, the selected halo at $z_i$  is disqualified as a candidate initial halo for DCBH formation. This process is repeated with increasing redshift until we identify a halo for which all progenitor haloes have virial temperatures below $10^4$\,K. Such a halo is considered a potential candidate for DCBH formation. 

Moreover, to further identify the potential DCBH host haloes, the cooling mechanism must be included to prevent gas fragmentation in these candidate haloes. Our criterion is that all progenitors of the candidate haloes must satisfy the following conditions:
\be
t_{\rm Hubble}<t_{\rm cool}\sim \frac{3nk_BT}{2\Lambda_{\rm eff}},
\ee
where $t_{\rm Hubble}$ is the Hubble time in the $\Lambda$CDM model, and this condition ensures that the halo gas does not lose significant energy via cooling on the Hubble timescale. The $n$ is the gas particle density given by:
\be
n\sim6f_{\rm gas}\left[\frac{T_{\rm vir}}{10^3\,{\rm K}}\right]^{3/2}{\rm cm}^{-3},
\ee 
with $f_{\rm gas}=0.2$ is a fitting parameter \citep{scoggins2024}. Here, $t_{\rm cool}$ is the cooling time, and $\Lambda_{\rm eff}$ is the effective cooling rate, which accounts for the competition between gas cooling and dynamical heating processes:
\be
\Lambda_{\rm eff}=n_{\rm H}n_{{\rm H}_2}\Lambda_{\rm cool}-\Gamma_{\rm heat}.
\ee
 The first term in the equation describes the cooling process  (with rate $\Lambda_{\rm cool}$  as discussed by \citet{HM1979}, \citet{GP2013}), which occurs due to the transition from atomic hydrogen to molecular hydrogen. This transition involves the joint reactions: ${\rm H}+{\rm e}^-\rightarrow{\rm H}^{-}+\hbar \nu$ and ${\rm H}+{\rm H}^{-}\rightarrow{\rm H}_2^{-}+{\rm e}^-$, and the $n_{\rm H},n_{{\rm H}_2}$ are the density of the atomic hydrogen and molecular hydrogen. Note that only a small fraction of atomic hydrogen transit to molecular hydrogen. Therefore, we express the atomic hydrogen density as  $n_{\rm H}=f_{\rm H}n\mu$ where the $f_{\rm H}=0.76, \mu=1.22$ are the mass fraction of hydrogen element, mean molecular mass, respectively. The molecular hydrogen density is 
\be
n_{{\rm H}_2}=k_9n_{\rm H} n_e/k_{\rm LW},
\ee
where $k_9$ characterises the reaction rate and can be found in \citet{Oh2002}, the $n_e=1.2\times 10^{-5}n_{\rm H}\sqrt{\Omega_m}/(\Omega_b h)$ is the post-recombination residue electron density. The $k_{\rm LW}$ is the dissociation rate of ${\rm H}_2$ by Lyman-Werner radiation, which will be elaborated later. The $\Gamma_{\rm heat}$ is the dynamical heating rate derived in \citep{yoshida2003}:
\be
\Gamma_{\rm heat}=\frac{n k_B}{\gamma-1}\frac{dT_{\rm vir}}{dt},
\ee
where the virial temperature can be related to the halo's virial mass via Eq.\,\eqref{eq:virial_temperature}.

Applying this criterion to the analysis of the merger tree allows us to complete the halo initialization procedure, which is designed to identify all  DCBH halo candidates where none of the progenitor haloes have undergone star formation. An additional point worth mentioning is the computation of LW radiation intensity. A shortcoming of the semi-analytical merger tree approach is the lack of spatial information on the haloes, and we follow the phenomenological model used in \citet{Dijkstra2008}, \citet{Li2021} and \citet{scoggins2024}, where the mean LW radiation received by a halo with mass $M_{\rm halo}$ at redshift $z$ is:
\be
\begin{split}
\bar {J}_{\rm LW}&(M_{\rm halo},z)=\\
&\int^{m_{\rm max}}_{m_{\rm min}}\int^{r_{\rm max}}_{r_{\rm min}}dmdr \frac{dN_{\rm halo}(m,r)}{dmdr}\frac{L_{\rm LW}}{16\pi^2 r^2}.
\end{split}
\ee
In this formula, the $L_{\rm LW}/4\pi r^2$ is the LW luminosity per unit area, which was given in \citet{Dijkstra2014}. The $dN_{\rm halo}(m,r)/dmdr$ is the population of haloes within the mass range $m\sim m+dm$ and the spherical shell of radius $r\sim r+dr$, which can be represented as:
\be
\begin{split}
&\frac{dN_{\rm halo}(m,r)}{dmdr}dm dr=\\
&\quad4\pi r^2dr(1+z)^3\frac{dn_{\rm halo}(m,z)}{dm}dm[1+\xi(M_{\rm halo},m,z,r)].
\end{split}
\ee
In the above formula, the spatial information is empirically encoded in the two-point halo correlation function $\xi(M_\mathrm{halo},m,z,r)$, representing the probability of finding a halo of mass $m$ at distance $r$ from a halo of mass $M_{\rm halo}$ \citep{Mo1996,SB2002,iliev2003}.

However, as noted by \citet{scoggins2024}, there will be stochastic variations of the surrounding haloes, which leads to a probability for a halo with mass $M_{\rm halo}$ at redshift $z$ to recieve Lyman-Werner radiation of $J_{\rm LW}(M_{\rm halo},z)$ as:
\be
P_{\rm halo}(x)=Ae^{-2|x-x_c|},
\ee
where $x={\rm log}_{10}[J_{\rm LW}(M_{\rm halo},z)]$ and $A$ is a normalization constant and $c\equiv{\rm log}_{10}[\bar{J}_{\rm LW}(M_{\rm halo},z)]$ \citep{lupi2021}.  For a candidate atomic cooling halo, we randomly sample a value $J_{\rm LW}$ by using the above $P(x)$, and then propagate the ratio $J_{\rm LW}/\bar{J}_{\rm LW}$ to all its progenitors with increasing redshift $z$. The physical picture behind this propagation of this ratio is that the candidate halo and all its progenitors should be exposed in a similar LW radiation environment.

The above calculations are driven by the assumption that MBHs form in haloes exposed to a supercritical LW flux.  This scenario naturally leads to a low DCBH number density and a low BH merger rate, consistent with results from numerous existing studies \citep{Dijkstra2014,Habouzit2016,Chiaki2023}. However, alternative approaches based on rapid halo growth and low metallicity can yield a significantly higher DCBH number density and merger rate, potentially representing an optimistic scenario for observations \citep{mcCaffrey2025}.

Another point that needs to be noticed is the self-shielding the LW radiation by the dense ${\rm H_2}$ column along the propagation line of the LW flux, which can reduce the total radiation received by the halo core. The self-shielding factor was extensively analyzed and modeled as a fitting formula by \citet{Wolcott-Green2017} and then used in \citet{scoggins2024}, of which the method is also adopted in our study\,(for details, see \citet{Wolcott-Green2017} and \citet{scoggins2024}).\\

\begin{figure}
    \centering
    \includegraphics[width=1.0\linewidth]{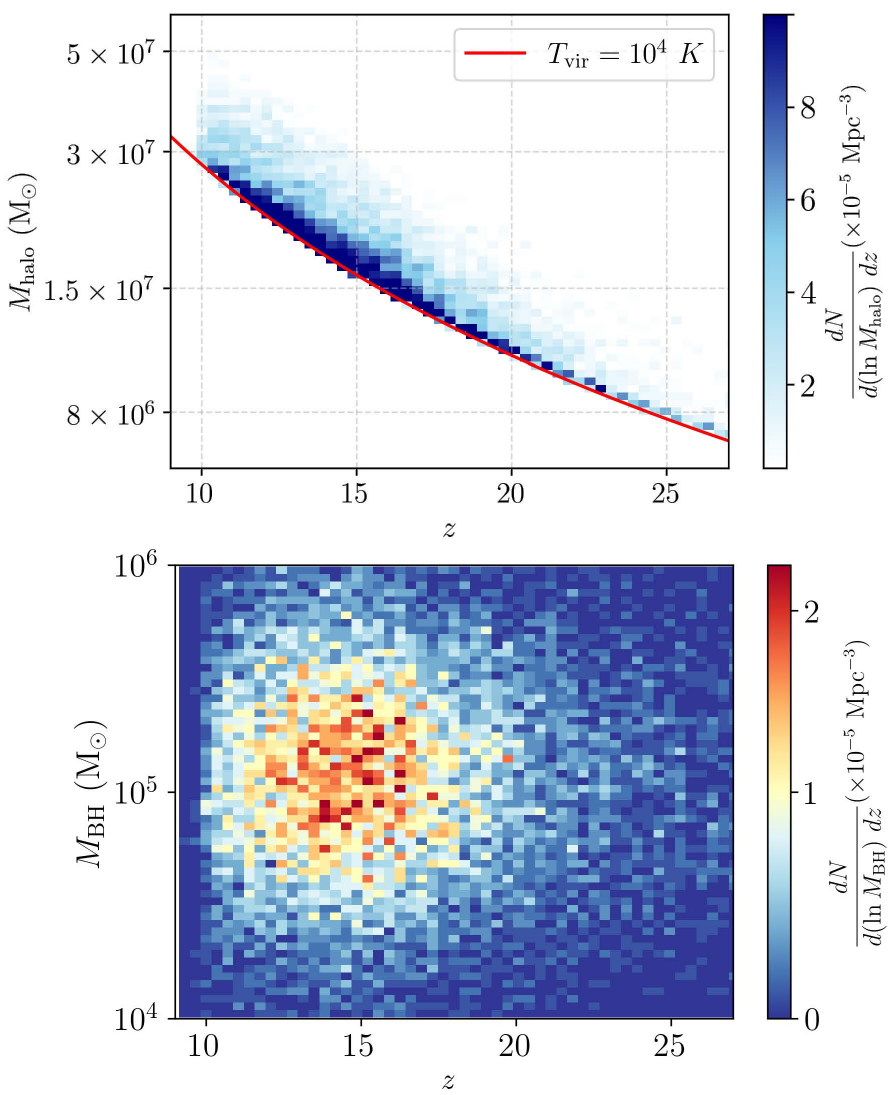}
    \caption{Upper panel: Mass and redshift\,($10<z<25$) distribution of the host dark matter halo of the DCBH identified in our merger tree. The red boundary exists due to the requirement of the Viral temperature $T_{\rm vir}>10^4$\,K. The halo mass is around $M_{\rm halo}\approx 2\times 10^7\,M_\odot$. Lower panel: Mass and redshift distribution of the direct-collapse black holes identified in our merger tree, it is clear that most of the DCBHs are generated during the cosmic dawn where $10<z<15$.}
        \label{fig:dcbh_halo_imf}
    \end{figure}

\subsubsection{Evolution of DCBHs in the Merger Tree}
The above discussion illustrates our method for obtaining the haloes where DCBHs site, and then we assign each halo with a seed BH by Monte-Carlo sampling following the initial mass function of DCBH,
\be
\frac{dn}{d{\rm ln}M_0}= \frac{1}{M_0\sigma_0\sqrt{2\pi}} \exp\left(-\frac{(\ln M_0 - \mu)^2}{2\sigma_0^2}\right),
\ee
where $\mu=11.7$ and $\sigma_0=1.0$ \citep{Ferrara2014,basu2019}. These DCBHs will evolve by gas accretion, merger, and interaction with the surrounding stars.  In our analysis we parameterize the gas accretion with rate:
\be\label{eq:gas_accretion_rate}
\dot{M}_{a}=\dot{M}_{\rm BH}=\frac{\eta_{\rm acc} L_{\rm Edd}}{\epsilon_\mathrm{r} c^2}=\frac{\eta_{\rm acc}}{\epsilon_\mathrm{r}}\frac{4\pi G \mu m_pM_{\rm BH}}{\sigma_{\rm T}c},
\ee
where the $\eta_{\rm acc}$ is the accretion efficiency, $\epsilon_\mathrm{r}$ is the radiative efficiency representing the conversion rate from the mass to the energy and the $L_{\rm Edd}$ is the Eddington luminosity \citep{Edd}. In this work, we assume a continuous and rapid gas supply with fiducial values $\epsilon_\mathrm{r} = 0.1$ \citep{haiman2001,madau_rees2001,volonteri2003,li2007}.
The merger of BH follows the merger of the dark matter haloes in the merger tree, where we do not consider the detailed evolution of binary BHs during the halo merger in this preliminary study. 
The contribution from the Pop-III TDEs to the evolution of the DCBH can be accounted for by integrating the model in Section\,\ref{sec:tde_modelling} with the merger tree, and the mass increasing rate given by:
\be
\dot{M}_{\rm TDE-III}\approx\langle m_*\rangle \Gamma_{\rm TDE}.
\ee
 \begin{figure}
        \centering
\includegraphics[width=1\linewidth]{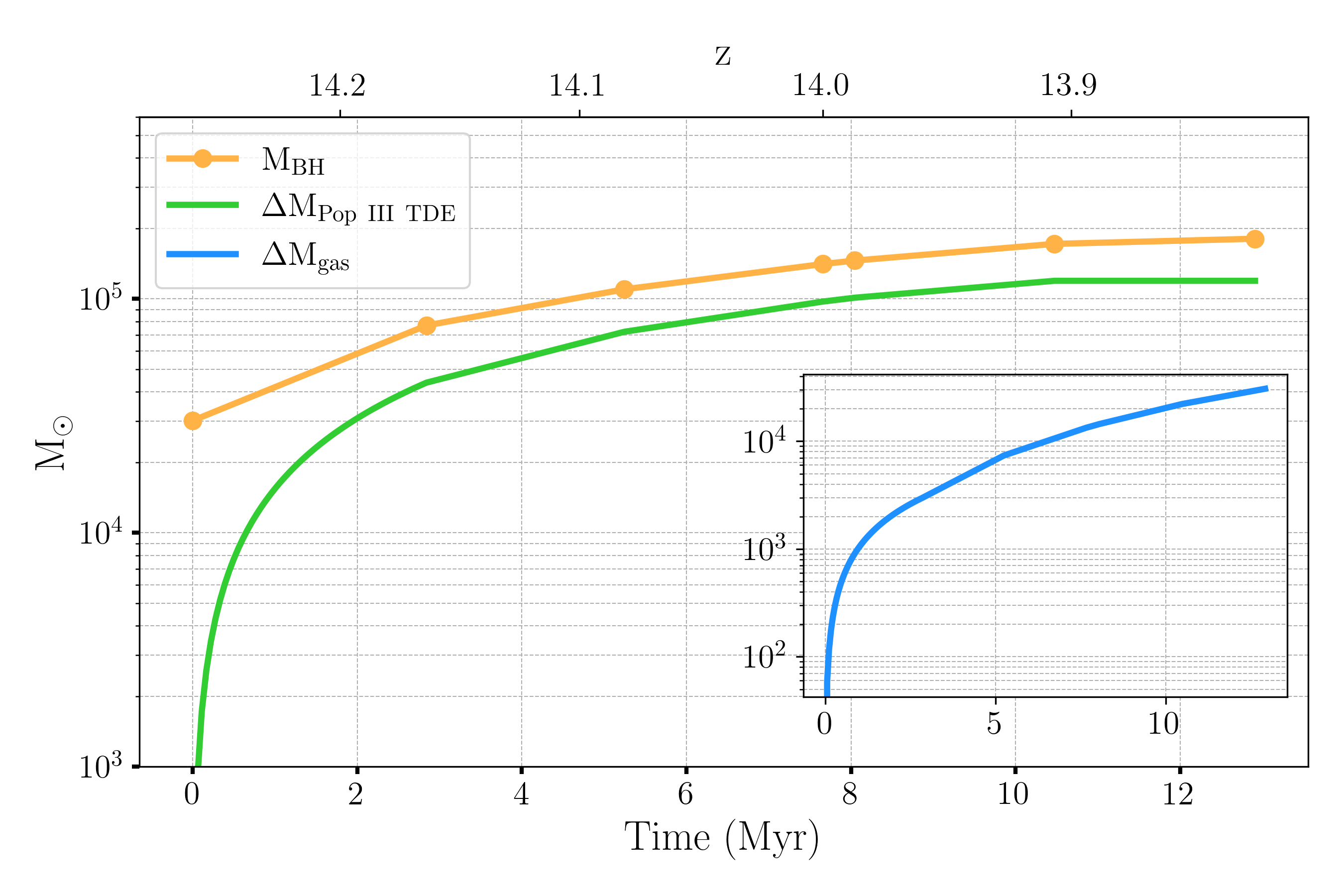}
        \caption{Mass growth of a $3\times10^4M_\odot$ DCBH due to the gas accretion and Pop III TDE before it undergoes a merger with another DCBH. The orange line represents the mass evolution of the DCBH, while the green and blue curves represent the contribution from Pop-III TDE and the gas accretion, respectively. The accretion rate is assumed to be at the Eddington limit.}
        \label{fig:one_bh}
    \end{figure}
In Figure\,\ref{fig:one_bh}, we present the mass growth of a $3\times 10^4\,M_\odot$ DCBH considering the above evolution channels, adopting Model-1 for Pop-III stars ($\langle m_*\rangle=70\,\mathrm{M_\odot},\,t_\mathrm{life}=10\,\mathrm{Myr}$) and our fiducial environmental parameters ($\gamma=7/4,\,\alpha=0.17,\,\varepsilon=0.1$). We find that Pop-III TDEs dominate the DCBH's mass growth during the first 5\,Myr, contributing nearly an order of magnitude more than gas accretion. While Model-2 yields qualitatively similar results, we employ Model-1 for all subsequent cosmological simulations.

\section{Results} \label{sec:results}
In this section, we investigate the impact of Pop-III TDEs on the cosmological evolution of MBHs. This is achieved by integrating the TDE modeling described earlier with a semi-analytical merger tree framework. We begin by examining the influence of Pop-III TDEs on the evolutionary pathways of MBHs using the fiducial model as a representative case. Subsequently, we average over the results across all merger trees to derive statistically robust conclusions that more accurately reflect realistic scenarios. Finally, we demonstrate the effects of Pop-III TDEs on the mass function of massive black holes across different redshifts, highlighting observational prospects at the cosmic dawn epoch.

\subsection{DCBH growth trajectory}
\begin{figure}
        \centering
        \includegraphics[width=1\linewidth]{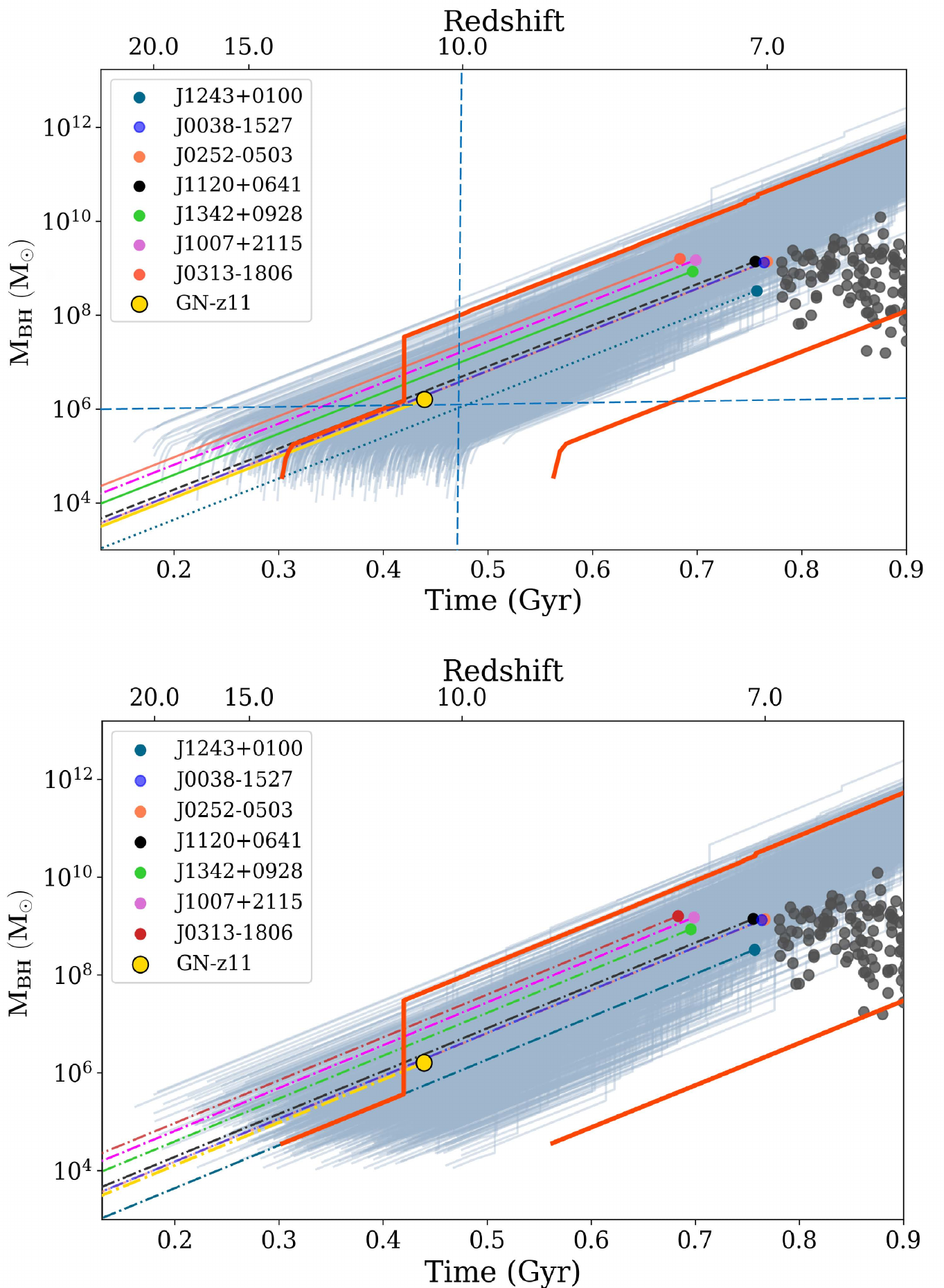}
        \caption{DCBH evolution trajectory. The light blue lines represent the main branches of each merger tree: with the Pop-III TDE contribution\,(upper) and without the Pop-III TDE contribution \,(lower). The gas accretion rate is set to be the Eddington limit.  The solid red lines denote the upper and lower limits of mass growth, starting with an initial black hole mass of $M_\mathrm{BH} < 5 \times 10^4 \mathrm{M_\odot}$ and evolving through Pop III TDEs at the onset. Colored circles indicate quasars observed at $z \geq 7$, as reported by \citet{mortlock2011}, \citet{banados2018}, \citet{matsuoka2019}, \citet{yang2020} and \citet{wang2018,wang2020,wang2021}, with their gas-accretion growth tracks shown as solid lines. Grey symbols represent black holes measured in quasars at $z \sim6- 7$ \citep{inayoshi2020}. Notably, the yellow circle with a black edge represents GN-z11, recently observed by JWST at a high redshift of $z = 10.6$ \citep{GNz11}.}
        \label{fig:all_tree}
    \end{figure}

Using the above methods, we have simulated the DCBH growth, including the Pop-III TDE channel, gas accretion, and MBH main-branch mergers. The resulting evolution trajectories in $10^3$\,merger trees are traced in Figure\,\ref{fig:all_tree} assuming the Eddington accretion rates $\dot{M}_a=\dot{M}_{\rm Edd}$ and the comparison to the scenarios without  Pop-III TDE events are also shown. In plotting these trajectories, we have used our fiducial model parameters listed in Table \ref{tab:physics_parameters}. The recently observed high-redshift quasars are also presented in Figure\,\ref{fig:all_tree} as the colored circles, together with their evolutionary trajectories assuming only the gas-accretion channel\,(named as gas-accretion trajectory in the following), as reported by \citet{wang2021}. 

    \begin{figure}
        \centering
        \includegraphics[width=1\linewidth]{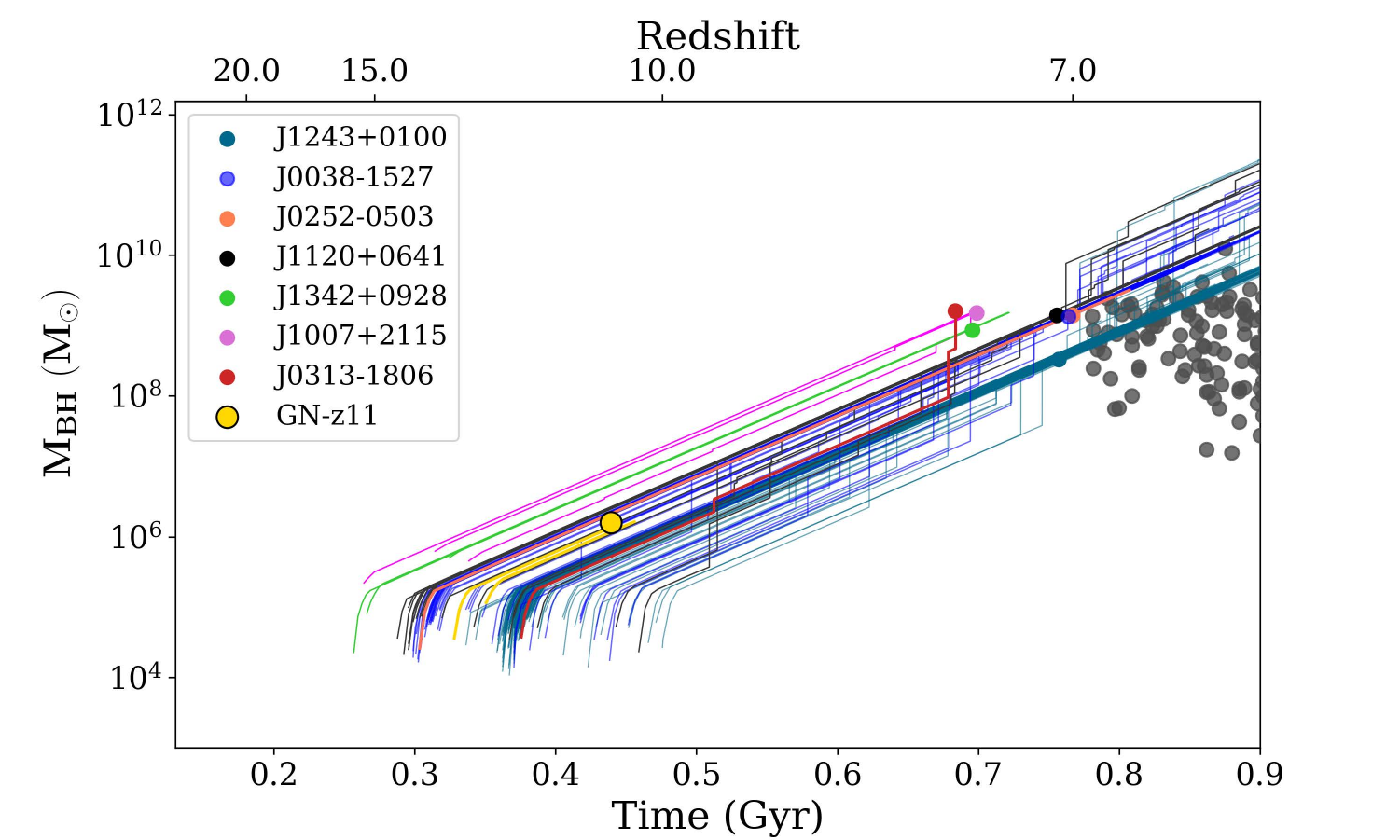}
        \caption{Potential trajectories for the UV-selected QSO  observed for $z>7$. We select those DCBH evolution trajectories that intersect with the circular source points. There can be multiple evolution paths corresponding to one high-redshift quasar.}
        \label{fig:intersections}
    \end{figure}

    \begin{figure}
        \centering
        \includegraphics[width=1\linewidth]{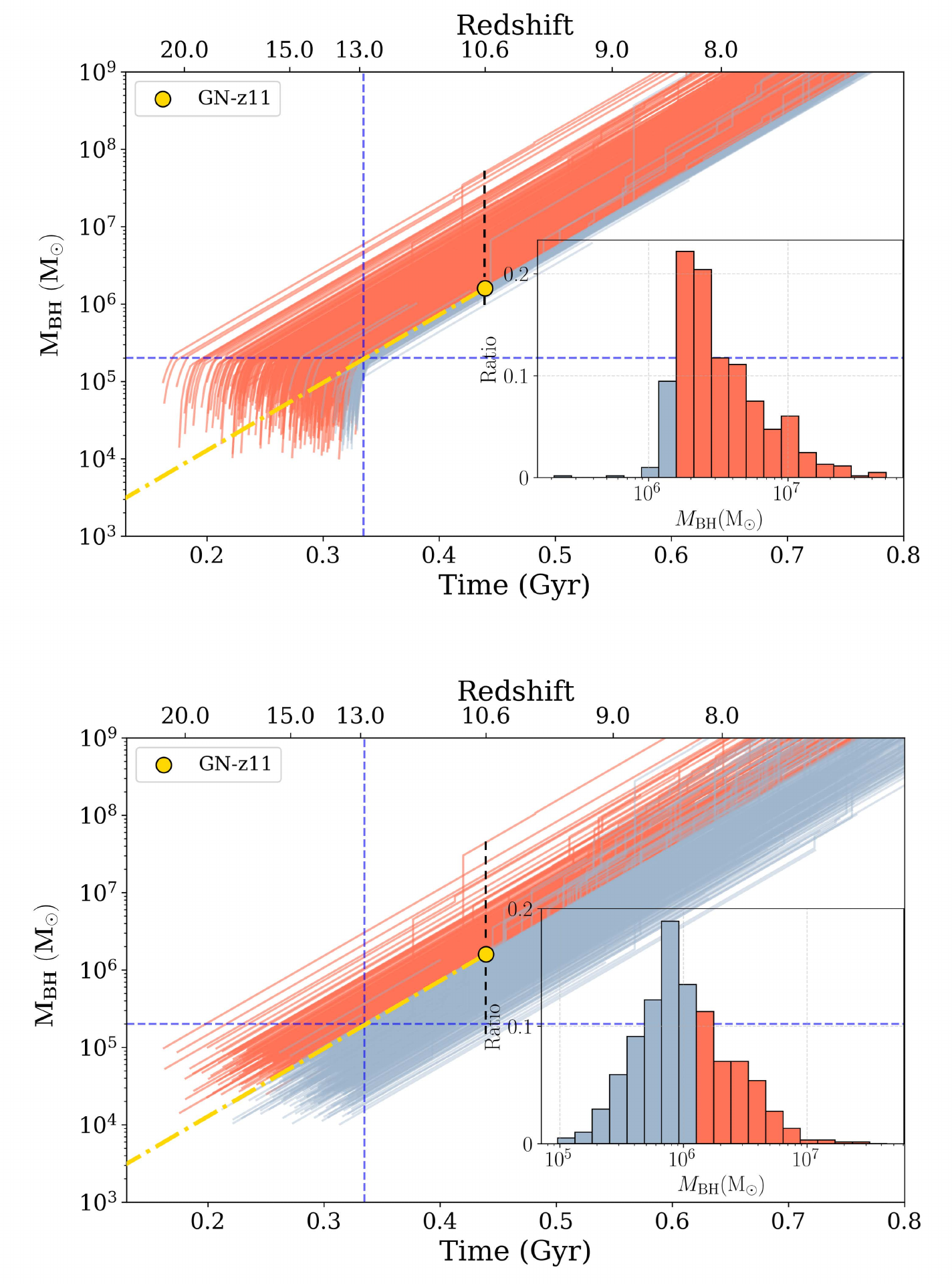}
        \caption{Evolution of DCBHs: A High-Redshift Perspective: We trace the growth of DCBHs in the early universe, using GN-z11 as a fiducial reference. The horizontal dashed line indicates the maximum mass attainable by a DCBH ($<10^5~\mathrm{M_\odot}$) through Pop-III TDEs. The upper panel shows that including Pop-III star TDEs enables 90\% of DCBHs with initial masses of $10^4-10^5\,\mathrm{M_\odot}$ to surpass GN-z11 in mass. In contrast, the lower panel demonstrates that without Pop-III TDEs, only $\sim$25\% of DCBHs grow above GN-z11’s mass.}
        \label{fig:zoom-in}
    \end{figure}

Compared to the gas-accretion trajectory, our trajectories are affected by both the merger and the Pop-III TDE. To explore the potential multiple evolution pathways of the $z>7$ UV-selected QSOs, we select a subset of DCBH evolution trajectories that intersect these sources, as illustrated in Figure\,\ref{fig:intersections}. Our primary focus is the impact of Pop-III TDEs. These results indicate the possibility that the DCBH's growth at the initial $\sim 5-10$\,Myr is strongly boosted by the Pop-III TDEs.  However, since the mass increase due to gas accretion is an exponential process as shown in Eq.\,\eqref{eq:gas_accretion_rate}, the contribution of gas accretion will overwhelm that of the Pop-III TDE contribution in the late evolutionary stage, which is consistent with the gas accretion trajectory. Furthermore, as redshift decreases, black hole mergers become more frequent and their contribution to mass growth eventually surpasses that of Pop-III TDEs as well. Therefore, to probe the contribution of Pop-III TDEs, high redshift universe $10<z<20$ must be probed. 

To assess the role of Pop-III TDEs in high-redshift sources like GN-z11, we analyze the evolution trajectories of DCBH initially formed at $z\geq 13$ with initial masses of $10^4-10^5\mathrm{M_\odot}$, and compare with the scenarios without Pop-III TDE contribution. As shown in Figure~\ref{fig:zoom-in}, our results demonstrate that when Pop-III TDEs are included, $90\%$ of these DCBHs grow to be heavier than GN-z11, whereas only $25\%$ do so without Pop-III TDEs. This suggests the possibility of a significant enhancement of the high-redshift AGN formation due to the contribution of the Pop-III TDEs.

\begin{figure}
        \centering
        \includegraphics[width=1.0\linewidth]{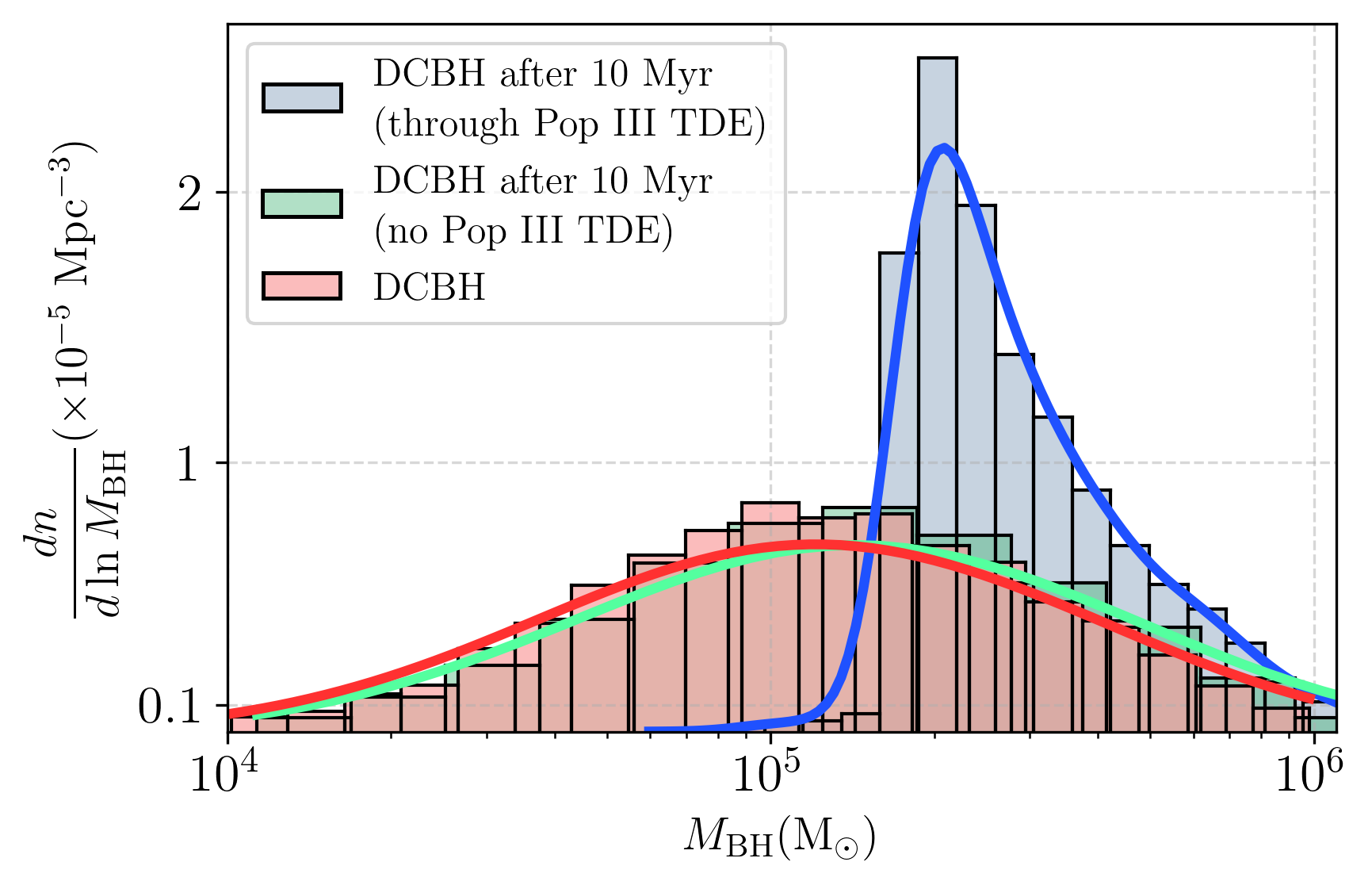}
        \caption{Three mass functions for direct-collapse black holes during the cosmic down $z\geq10$: the initial mass function, and two evolved mass functions after $10^7$ years\,(corresponding to an incremental redshift $\Delta z\sim 0.1-1$), distinguished by whether Pop-III TDEs are included or excluded.}
        \label{fig:dcbh_IMF}
    \end{figure}

\begin{figure}
    \centering
    \includegraphics[width=\linewidth]{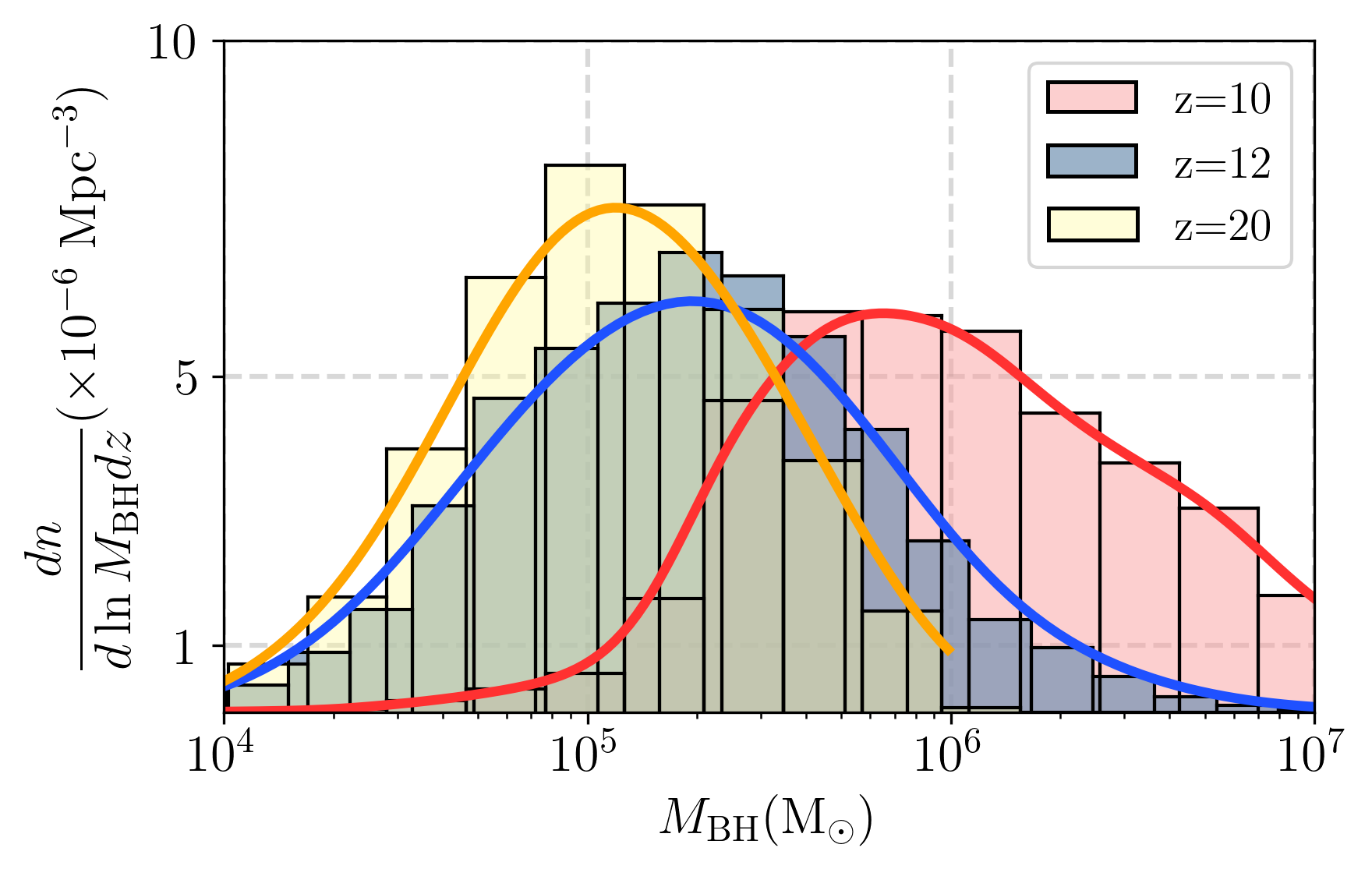}
        \includegraphics[width=\linewidth]{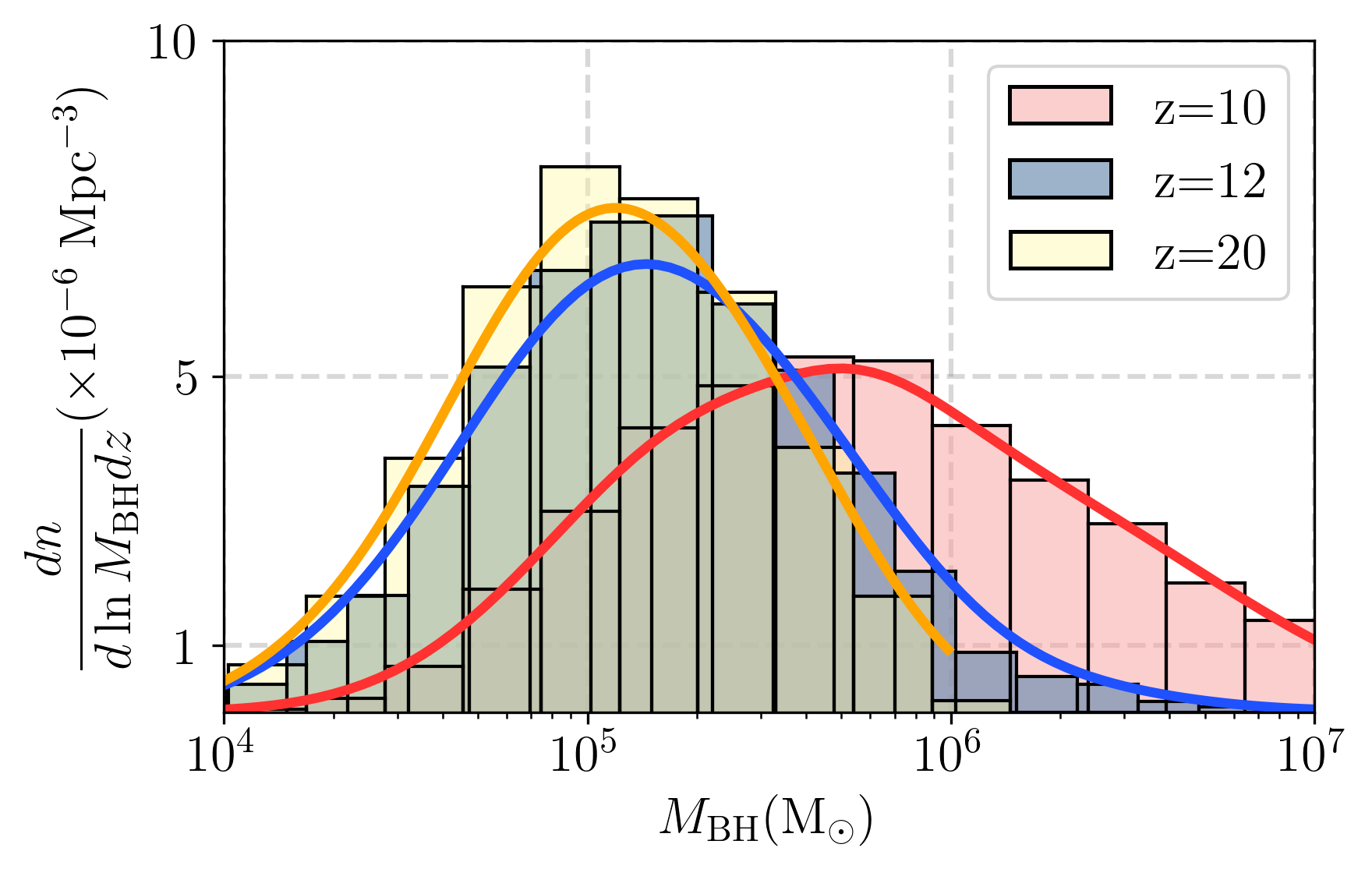}
    \caption{Evolution of black hole mass functions at different redshifts during the cosmic dawn $z \geq 10$. Upper panel: we include the contribution of gas accretion, Pop III TDEs and also the MBH merger events. Lower panel: the Pop-III TDE contribution is excluded for comparison. The mass distribution is significantly modified with the contribution of Pop-III TDEs.}
    \label{fig:BHMFz}
\end{figure}

Now let us take a closer look at the DCBH distributions generated by our simulation during the cosmic dawn,  represented by  Figure\,\ref{fig:dcbh_IMF} and Figure\,\ref{fig:BHMFz}, which are obtained by slicing the Figure\,\ref{fig:all_tree} using the long-dashed lines. Counting all the DCBH populations on a BH mass slice\,(horizontal long-dashed line) and then scan over the BH mass,  we have the DCBH mass distribution function shown in Figure\,\ref{fig:dcbh_IMF} 
\be
\frac{dn_{\rm BH}(M_{\rm BH})}{d{\rm ln}M_{\rm BH}}=\int dz \frac{dn_{\rm BH}(M_{\rm BH}, z)}{dz\times d{\rm ln}M_{\rm BH}},
\ee
which is significantly evolved after 10\,Myr, due to the Pop-III TDEs. In particular, those DCBHs with initial masses $\sim 10^4\,M_\odot$ can quickly increase their masses to above $10^5\,M_\odot$ and distorted the DCBH initial mass function, while the gas accretion during the initial 10\,Myr only minorly contributes to the change of the distribution function.  Moreover, during the first 10\,Myrs, most of the DCBH will not undergo binary merger.

We also present the cosmological evolution of DCBH mass-redshift distribution $dn(M_{\rm BH}, z)/(dz\times d{\rm ln}M_{\rm BH})$ shown in Figure\,\ref{fig:BHMFz}, which is obtained by counting the mass population at redshift $z=20,12,10$. This mass-redshift distribution and its evolution at the high-redshift universe, in principle, carry the imprint of the role of Pop-III TDE on the DCBH growth, and has the potential to be probed by the future observations\,(see\,Sec.\,\ref{sec:observation}).

\subsection{Observational probe}\label{sec:observation}
Gravitational wave\,(GW) radiated by the massive binary black hole system can be detected by future spaceborne gravitational wave detectors, such as LISA/Tianqin/Taiji \citep{LISA,taiji,tianqin}. For instance, the signal-to-noise ratio for the LISA detector can reach $\sim 100$ for $10^5\,M_\odot$ MBHs at $z=15$, and can measure the mass/spin parameters of coalescing MBHs to $10^{-2}$ accuracy in $10^4-10^7$ Massive BBHs up to $z\sim20$ \citep{LISAWB2019}. Therefore, probing the properties of the seed BHs, including DCBH as the heavy seed BHs, is one of the important science cases for spaceborne GW detectors.  For example, if there is sufficient DCBH merger events in the high redshift universe, detection of the DCBH merger events could provide information on the DCBH mass distributions, hence providing an observational probe to the DCBH growth channel driven by Pop-III TDEs.

\begin{figure}
    \centering
\includegraphics[width=1.0\linewidth]{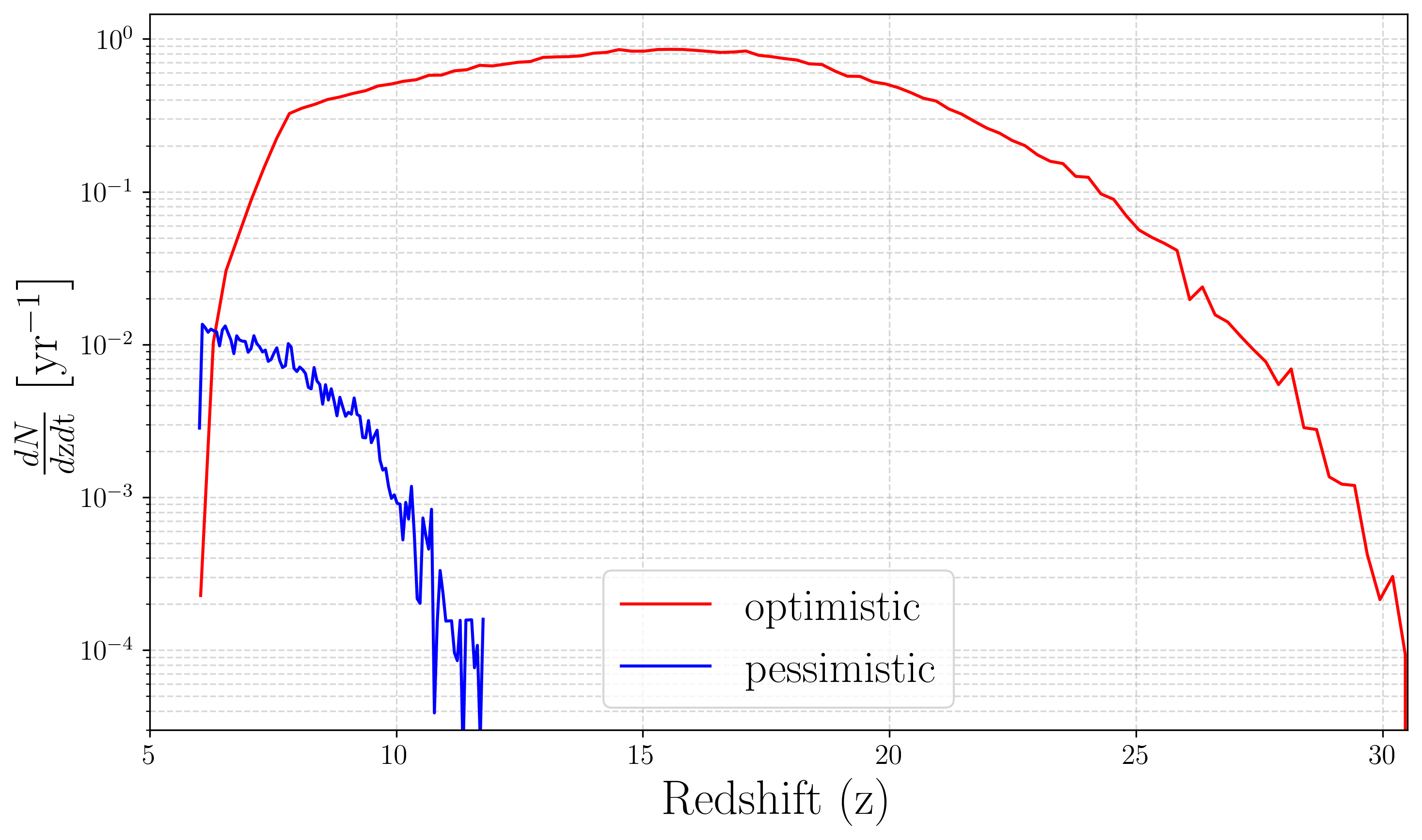}
\includegraphics[width=1.0\linewidth]{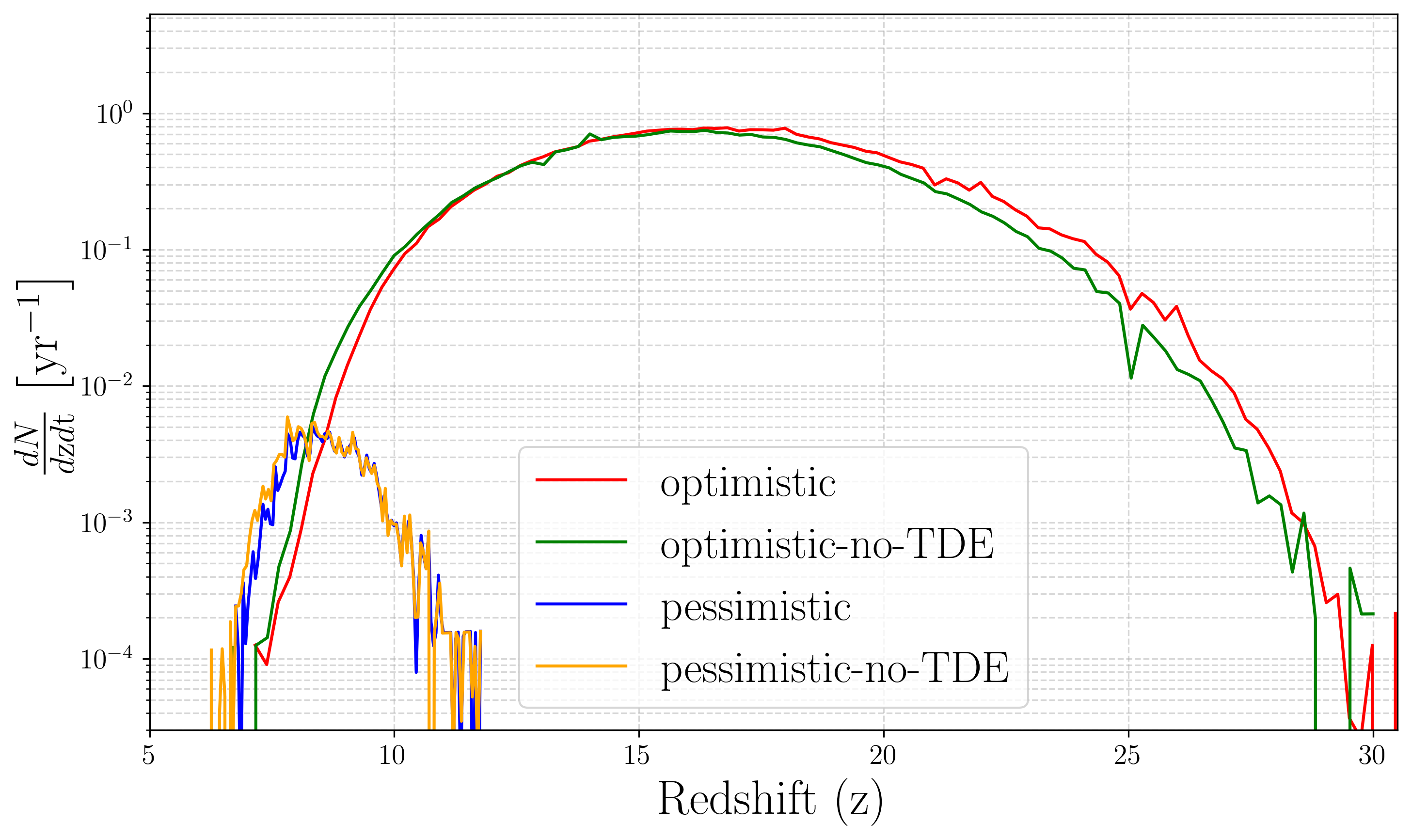}
    \caption{Upper panel: Redshift evolution of the DCBH merger rate, showing optimistic and pessimistic estimates under different host halo selection criteria --- rapid gas inflow (optimistic) versus strong Lyman-Werner flux (pessimistic).  Lower panel: Redshift evolution of DCBH merger rate detectable as GW event rates comparing scenarios with and without Pop-III TDEs. The inclusion of Pop-III TDEs shows only a marginal impact on the merger rate.}
    \label{fig:merger_rate_1}
\end{figure}

However, large uncertainties still exist in estimating the MBH merger rate at the high redshift universe, which will directly affect our evaluation of the detection rates.  For example, there are optimistic estimations by \citet{RN18a} showing that $\sim 20-100$ heavy seed merger events for a four-year observation. In contrast, the pessimistic estimations by \emph{Astrid} simulations predicted a much lower event rate $\sim 10^{-2}-10^{-3}$\,$\mathrm{yr^{-1}}$ \citep{degraf2024}. Our merger tree simulations—which adopt relatively strict DCBH criteria—also align with the more pessimistic scenario, as illustrated in Figure\,\ref{fig:merger_rate_1}. Such a large uncertainty reflects the uncertainties of the environmental modeling of the high-redshift universe, for instance, the treatment of metallicity in the DCBH selection criterion differs among the above results \citep{mcCaffrey2025}. 

Moreover, adjusting the parameters for LW-radiation background and selecting rapidly growing haloes in our model can also yield an optimistic estimation of merger events aligned with \citet{mcCaffrey2025}, as shown in the upper panel of Figure\,\ref{fig:merger_rate_1} for comparison. 

The GW event rate of DCBH mergers is shown in the lower panel of Figure\,\ref{fig:merger_rate_1}, presenting both pessimistic and optimistic estimates. To investigate the Pop-III TDE growth channel, we compare scenarios with and without Pop-III TDEs for each case. Our results indicate that Pop-III TDEs have only a marginal impact on the overall DCBH merger rate. However, if future GW observations align with the optimistic scenario ($1 ~\mathrm{yr^{-1}}$ peaking around $z\approx15$), the mass functions of the MBHs across different redshifts could be precisely measured and determined as illustrated in Figure\,\ref{fig:BHMFz}. This would place crucial constraints on the rapid growth mechanisms operating in the early evolutionary stages of DCBHs.

Detecting Pop-III TDEs through optical and infrared observations presents a promising avenue for understanding early black hole growth. Recent studies suggest that Pop-III TDEs exhibit higher mass fallback rates and longer evolution timescales compared to those involving solar-type stars, resulting in luminous flares that are redshifted into the infrared spectrum \citep{karmen2025}. Instruments like the James Webb Space Telescope (JWST) and the upcoming Nancy Grace Roman Space Telescope are well-suited to detect these infrared signatures, potentially observing several such events per year \citep{Rigby_2023,spergel2015}. These observations could provide indirect evidence of Pop-III stars and offer insights into the early stages of supermassive black hole formation.

\section{Conclusion and discussion}\label{sec:conclusion}
In this work, using the semi-analytical method, we have explored the TDEs of Pop-III stars by DCBHs as a new possible growth channel of massive black holes during the cosmic dawn. Whereas the dominant DCBH growth channel at $z<10$ is the gas accretion and the massive black hole mergers, our analysis shows that the Pop-III TDEs can significantly boost the growth of the DCBHs during their first 10\,Myr evolution. This process enhances the formation of high redshift QSO/AGN, and modifies the mass distribution of the DCBHs at the cosmic dawn.  In the following, several comments are made on this work from both the methodological aspect, and the future science opportunities.

The general physical process can be summarised as follows. At the cosmic dawn, the cold and metal-free baryonic gas transports through the cosmic filament ``veins" and flows into the dark matter halo center located at the high-density node of multiple filaments, with a deep gravitational potential.  The baryonic flow then turns to a central accretion disk surrounding the DCBH, and Pop-III stars are born within the fragmentation radius. Through gravitational scattering, these stars then form a nuclear star cluster, and the dynamical equilibrium population $\sim 100-200$ can be reached by balancing the star supply and the star consumption by TDEs. Within the lifetime of  Pop-III stars\,(say 10\,Myr), the DCBH growth is dominated by the Pop-III TDEs over the gas accretion with the Eddington accretion rate.  In this work, the phenomenological modeling for the Pop-III TDEs around DCBH can also be elaborated in the future. For instance, the statistical method is used when the equilibrium star population in the NSC is $100-200$, which needs to be further checked using N-body simulations;  The regulation by ultraviolet radiation also needs a more careful study.\\

Our analysis also evolves the DCBHs in the cosmological merger tree, considering the effect of Pop-III TDEs. We grow $10^3$ merger trees from $z=6$ to $z=30$. The Lyman-Werner radiation background and the cooling criterion are applied to the merger tree to select DCBH host haloes. The DCBH is distributed randomly into these haloes following a standard initial mass function. Taking into account the growth channels of MBH merger, Pop-III TDE, and gas accretion, the evolution trajectories of the DCBHs are obtained, which are our main results shown in Figure\,\ref{fig:all_tree}. Although the merger-tree-generated dark-matter halo distribution agrees well with that obtained by the N-body simulation at the high redshift universe \citep{parkinson2008}, our preliminary method for identifying the DCBH host halo and the BH merger likely ignored many detailed physical factors that may affect the evolution history. For instance, our treatment of the Lyman-Werner radiation is purely phenomenological, since the merger tree cannot provide the spatial correlation information of the haloes. When DCBH is formed, it can wander in the halo potential well and sink into the center in a delayed way, which we did not take into account in our simulation. We also ignored the dynamical details of the MBH inspiralling and the possibility that the merger remnants can be kicked out of the halo. Instead, we assume that the DCBH is always located at the halo center and merges simultaneously with the halo merger. Moreover, SMS formation would generate an H-II region at the halo center that subsequently suppresses cold accretion \citep{K23,masaki2024}.  This potential delay in DCBH growth warrants further investigation to quantify its impact on early black hole formation scenarios. Estimates of the influence of these simplifications call for a more detailed simulation.

Despite these imperfections of our preliminary modeling, Pop-III TDEs, as a potential rapid growth channel for DCBHs, could provide valuable insights for future observations. At high redshifts, MBH merger events detected by spaceborne GW detectors may shed new light on black hole formation and allow us to probe the early growth of black holes by examining the mass function and number density of DCBHs, thereby testing evolutionary scenarios driven by Pop-III TDEs.

\section{Acknowledgements}
This work is inspired by the conversation between Professor Yanbei Chen and Y.M. on the vision of the high-redshift universe five years ago. The authors also thank Professor John Wise for helpful discussions.  In addition, Y.\,M. and Z.\,W. would like to thank Mengye Wang for helpful discussions on TDE.  Y. M. would like to thank Xuefei Gong and Yan Wang for the discussion on the LISA merger rate. The research of Y.\,M., Z.\,W., C.\,W. and Y.\,L. is supported by the start-up funding provided by Huazhong University of Science and Technology\,(HUST), and in particular, the High-Performance Computing Platform ``Gravity-1" provided by the National Gravitation Laboratory. Y.\,L. is also supported by the Undergraduate Research funding No.\,1240026 provided by HUST. Z.C. is supported by National Key R\&D Program of China (grant no. 2023YFA1605600). Q.W.is supported by the National Natural Science Foundation of
China (grants 123B2041 and 12233007).

\bibliography{sample631}{}

\begin{thebibliography}{}
\expandafter\ifx\csname natexlab\endcsname\relax\def\natexlab#1{#1}\fi
\providecommand{\url}[1]{\href{#1}{#1}}
\providecommand{\dodoi}[1]{doi:~\href{http://doi.org/#1}{\nolinkurl{#1}}}
\providecommand{\doeprint}[1]{\href{http://ascl.net/#1}{\nolinkurl{http://ascl.net/#1}}}
\providecommand{\doarXiv}[1]{\href{https://arxiv.org/abs/#1}{\nolinkurl{https://arxiv.org/abs/#1}}}

\bibitem[{{Abac} {et~al.}(2025){Abac}, {Abramo}, {Albanesi}, {Albertini},
  {Agapito}, {Agathos}, {Albertus}, {Andersson}, {Andrade}, {Andreoni},
  {Angeloni}, {Antonelli}, {Antoniadis}, {Antonini}, {Arca Sedda}, {Artale},
  {Ascenzi}, {Auclair}, {Bachetti}, {Badger}, {Banerjee}, {Barba-Gonz{\'a}lez},
  {Barta}, {Bartolo}, {Bauswein}, {Begnoni}, {Beirnaert}, {Bejger}, {Belgacem},
  {Bellomo}, {Bernard}, {Grazia Bernardini}, {Bernuzzi}, {Berry}, {Berti},
  {Bertone}, {Bettoni}, {Bezares}, {Bhagwat}, {Bisero}, {Bizouard},
  {Blanco-Pillado}, {Blasi}, {Bonino}, {Borghese}, {Borghi}, {Borhanian},
  {Bortolas}, {Botticella}, {Branchesi}, {Breschi}, {Brito}, {Brocato},
  {Broekgaarden}, {Bulik}, {Buonanno}, {Burgio}, {Burrows}, {Calcagni},
  {Canevarolo}, {Cappellaro}, {Capurri}, {Carbone}, {Casadio}, {Cayuso},
  {Cerd{\'a}-Dur{\'a}n}, {Char}, {Chaty}, {Chiarusi}, {Chruslinska}, {Cireddu},
  {Cole}, {Colombo}, {Colpi}, {Comp{\`e}re}, {Contaldi}, {Corman},
  {Crescimbeni}, {Cristallo}, {Cuoco}, {Cusin}, {Dal Canton}, {D{\'a}lya},
  {D'Avanzo}, {Davari}, {De Luca}, {De Renzis}, {Della Valle}, {Del Pozzo}, {De
  Santi}, {Ludovico De Santis}, {Dietrich}, {Dimastrogiovanni}, {Domenech},
  {Doneva}, {Drago}, {Dupletsa}, {Duval}, {Dvorkin}, {Elias-Rosa}, {Fairhurst},
  {Fantina}, {Fasiello}, {Fays}, {Fender}, {Fischer}, {Foucart}, {Fragos},
  {Foffa}, {Franciolini}, {Gair}, {Gamba}, {Garcia-Bellido},
  {Garc{\'\i}a-Quir{\'o}s}, {{\'A}rp{\'a}d Gergely}, {Ghirlanda}, {Ghosh},
  {Giacomazzo}, {Gittins}, {Giudice}, {Goncharov}, {Gonzalez}, {Gori{\'e}ly},
  {Graziani}, {Greco}, {Gualtieri}, {Guidi}, {Gupta}, {Haney}, {Hannam},
  {Harms}, {Harutyunyan}, {Haskell}, {Haungs}, {Hazra}, {Hemming}, {Heng},
  {Hinderer}, {van der Horst}, {Hu}, {Husa}, {Iacovelli}, {Illuminati},
  {Inguglia}, {Izquierdo Villalba}, {Janquart}, {Janssens}, {Jenkins}, {Jones},
  {Kacskovics}, {Klessen}, {Kokkotas}, {Kuan}, {Kumar}, {Kuroyanagi}, {Laghi},
  {Lamberts}, {Lambiase}, {Larrouturou}, {Leaci}, {Lenzi}, {Levan}, {Li}, {Li},
  {Liang}, {Limongi}, {Liu}, {Llanes-Estrada}, {Loffredo}, {Long}, {Lope-Oter},
  {Lukes-Gerakopoulos}, {Maggio}, {Maggiore}, {Mancarella}, {Mapelli},
  {Marchant}, {Margiotta}, {Mariotti}, {Marriott-Best}, {Marsat},
  {Mart{\'\i}nez-Pinedo}, {Maselli}, {Mastrogiovanni}, {Matos}, {Melandri},
  {Mendes}, {Mendon{\c{c}}a Soares de Souza}, {Mentasti}, {Mezcua},
  {M{\"o}sta}, {Mondal}, {Moresco}, {Mukherjee}, {Muttoni}, {Nagar}, {Narola},
  {Nava}, {Navarro Moreno}, \& {Nelemans}}]{ET}
{Abac}, A., {Abramo}, R., {Albanesi}, S., {et~al.} 2025, arXiv e-prints,
  arXiv:2503.12263, \dodoi{10.48550/arXiv.2503.12263}

\bibitem[{{Alexander} \& {Bar-Or}(2017)}]{alexander2017}
{Alexander}, T., \& {Bar-Or}, B. 2017, Nature Astronomy, 1, 0147,
  \dodoi{10.1038/s41550-017-0147}

\bibitem[{{Alexander} \& {Hopman}(2009)}]{alexander2009}
{Alexander}, T., \& {Hopman}, C. 2009, \apj, 697, 1861,
  \dodoi{10.1088/0004-637X/697/2/1861}

\bibitem[{{Amaro-Seoane} {et~al.}(2017){Amaro-Seoane}, {Audley}, {Babak},
  {Baker}, {Barausse}, {Bender}, {Berti}, {Binetruy}, {Born}, {Bortoluzzi},
  {Camp}, {Caprini}, {Cardoso}, {Colpi}, {Conklin}, {Cornish}, {Cutler},
  {Danzmann}, {Dolesi}, {Ferraioli}, {Ferroni}, {Fitzsimons}, {Gair}, {Gesa
  Bote}, {Giardini}, {Gibert}, {Grimani}, {Halloin}, {Heinzel}, {Hertog},
  {Hewitson}, {Holley-Bockelmann}, {Hollington}, {Hueller}, {Inchauspe},
  {Jetzer}, {Karnesis}, {Killow}, {Klein}, {Klipstein}, {Korsakova}, {Larson},
  {Livas}, {Lloro}, {Man}, {Mance}, {Martino}, {Mateos}, {McKenzie},
  {McWilliams}, {Miller}, {Mueller}, {Nardini}, {Nelemans}, {Nofrarias},
  {Petiteau}, {Pivato}, {Plagnol}, {Porter}, {Reiche}, {Robertson},
  {Robertson}, {Rossi}, {Russano}, {Schutz}, {Sesana}, {Shoemaker}, {Slutsky},
  {Sopuerta}, {Sumner}, {Tamanini}, {Thorpe}, {Troebs}, {Vallisneri},
  {Vecchio}, {Vetrugno}, {Vitale}, {Volonteri}, {Wanner}, {Ward}, {Wass},
  {Weber}, {Ziemer}, \& {Zweifel}}]{LISA}
{Amaro-Seoane}, P., {Audley}, H., {Babak}, S., {et~al.} 2017, arXiv e-prints,
  arXiv:1702.00786.
\newblock \doarXiv{1702.00786}

\bibitem[{{Andalman} {et~al.}(2024){Andalman}, {Teyssier}, \&
  {Dekel}}]{Andalman2024}
{Andalman}, Z.~L., {Teyssier}, R., \& {Dekel}, A. 2024, arXiv e-prints,
  arXiv:2410.20530, \dodoi{10.48550/arXiv.2410.20530}

\bibitem[{{Ba{\~n}ados} {et~al.}(2018){Ba{\~n}ados}, {Venemans},
  {Mazzucchelli}, {Farina}, {Walter}, {Wang}, {Decarli}, {Stern}, {Fan},
  {Davies}, {Hennawi}, {Simcoe}, {Turner}, {Rix}, {Yang}, {Kelson}, {Rudie}, \&
  {Winters}}]{banados2018}
{Ba{\~n}ados}, E., {Venemans}, B.~P., {Mazzucchelli}, C., {et~al.} 2018, \nat,
  553, 473, \dodoi{10.1038/nature25180}

\bibitem[{{Barkana} \& {Loeb}(2001)}]{barkana_loeb2001}
{Barkana}, R., \& {Loeb}, A. 2001, \physrep, 349, 125,
  \dodoi{10.1016/S0370-1573(01)00019-9}

\bibitem[{{Barrow} {et~al.}(2018){Barrow}, {Aykutalp}, \& {Wise}}]{barrow2018}
{Barrow}, K. S.~S., {Aykutalp}, A., \& {Wise}, J.~H. 2018, Nature Astronomy, 2,
  987, \dodoi{10.1038/s41550-018-0569-y}

\bibitem[{{Basu} \& {Das}(2019)}]{basu2019}
{Basu}, S., \& {Das}, A. 2019, \apjl, 879, L3, \dodoi{10.3847/2041-8213/ab2646}

\bibitem[{{Begelman} {et~al.}(2008){Begelman}, {Rossi}, \&
  {Armitage}}]{begelman2008}
{Begelman}, M.~C., {Rossi}, E.~M., \& {Armitage}, P.~J. 2008, \mnras, 387,
  1649, \dodoi{10.1111/j.1365-2966.2008.13344.x}

\bibitem[{{Begelman} {et~al.}(2006){Begelman}, {Volonteri}, \&
  {Rees}}]{BVR2006}
{Begelman}, M.~C., {Volonteri}, M., \& {Rees}, M.~J. 2006, \mnras, 370, 289,
  \dodoi{10.1111/j.1365-2966.2006.10467.x}

\bibitem[{{Bond} {et~al.}(1991){Bond}, {Cole}, {Efstathiou}, \&
  {Kaiser}}]{bond1991}
{Bond}, J.~R., {Cole}, S., {Efstathiou}, G., \& {Kaiser}, N. 1991, \apj, 379,
  440, \dodoi{10.1086/170520}

\bibitem[{{Bower}(1991)}]{bower1991}
{Bower}, R.~G. 1991, \mnras, 248, 332, \dodoi{10.1093/mnras/248.2.332}

\bibitem[{{Bromm} \& {Loeb}(2003)}]{bromm_loeb2003}
{Bromm}, V., \& {Loeb}, A. 2003, \apj, 596, 34, \dodoi{10.1086/377529}

\bibitem[{{Bunker} {et~al.}(2023){Bunker}, {Saxena}, {Cameron}, {Willott},
  {Curtis-Lake}, {Jakobsen}, {Carniani}, {Smit}, {Maiolino}, {Witstok},
  {Curti}, {D'Eugenio}, {Jones}, {Ferruit}, {Arribas}, {Charlot}, {Chevallard},
  {Giardino}, {de Graaff}, {Looser}, {L{\"u}tzgendorf}, {Maseda}, {Rawle},
  {Rix}, {Del Pino}, {Alberts}, {Egami}, {Eisenstein}, {Endsley}, {Hainline},
  {Hausen}, {Johnson}, {Rieke}, {Rieke}, {Robertson}, {Shivaei}, {Stark},
  {Sun}, {Tacchella}, {Tang}, {Williams}, {Willmer}, {Baker}, {Baum},
  {Bhatawdekar}, {Bowler}, {Boyett}, {Chen}, {Circosta}, {Helton}, {Ji},
  {Kumari}, {Lyu}, {Nelson}, {Parlanti}, {Perna}, {Sandles}, {Scholtz},
  {Suess}, {Topping}, {{\"U}bler}, {Wallace}, \& {Whitler}}]{GNz11}
{Bunker}, A.~J., {Saxena}, A., {Cameron}, A.~J., {et~al.} 2023, \aap, 677, A88,
  \dodoi{10.1051/0004-6361/202346159}

\bibitem[{{Chandrasekhar}(1943)}]{ch1943}
{Chandrasekhar}, S. 1943, Reviews of Modern Physics, 15, 1,
  \dodoi{10.1103/RevModPhys.15.1}

\bibitem[{{Chiaki} {et~al.}(2023){Chiaki}, {Chon}, {Omukai}, {Trinca},
  {Schneider}, \& {Valiante}}]{Chiaki2023}
{Chiaki}, G., {Chon}, S., {Omukai}, K., {et~al.} 2023, \mnras, 521, 2845,
  \dodoi{10.1093/mnras/stad689}

\bibitem[{{Cole} {et~al.}(2000){Cole}, {Lacey}, {Baugh}, \& {Frenk}}]{cole2000}
{Cole}, S., {Lacey}, C.~G., {Baugh}, C.~M., \& {Frenk}, C.~S. 2000, \mnras,
  319, 168, \dodoi{10.1046/j.1365-8711.2000.03879.x}

\bibitem[{{Colpi} {et~al.}(2019){Colpi}, {Holley-Bockelmann}, {Bogdanovic},
  {Natarajan}, {Bellovary}, {Sesana}, {Tremmel}, {Schnittman}, {Comerford},
  {Barausse}, {Berti}, {Volonteri}, {Khan}, {McWilliams}, {Burke-Spolaor},
  {Hazboun}, {Conklin}, {Mueller}, \& {Larson}}]{LISAWB2019}
{Colpi}, M., {Holley-Bockelmann}, K., {Bogdanovic}, T., {et~al.} 2019, arXiv
  e-prints, arXiv:1903.06867, \dodoi{10.48550/arXiv.1903.06867}

\bibitem[{{DeGraf} {et~al.}(2024){DeGraf}, {Chen}, {Ni}, {Di Matteo}, {Bird},
  {Tremmel}, \& {Croft}}]{degraf2024}
{DeGraf}, C., {Chen}, N., {Ni}, Y., {et~al.} 2024, \mnras, 527, 11766,
  \dodoi{10.1093/mnras/stad3084}

\bibitem[{{Di Matteo} {et~al.}(2012){Di Matteo}, {Khandai}, {DeGraf}, {Feng},
  {Croft}, {Lopez}, \& {Springel}}]{DiMatteo2012}
{Di Matteo}, T., {Khandai}, N., {DeGraf}, C., {et~al.} 2012, \apjl, 745, L29,
  \dodoi{10.1088/2041-8205/745/2/L29}

\bibitem[{{Dijkstra} {et~al.}(2014){Dijkstra}, {Ferrara}, \&
  {Mesinger}}]{Dijkstra2014}
{Dijkstra}, M., {Ferrara}, A., \& {Mesinger}, A. 2014, \mnras, 442, 2036,
  \dodoi{10.1093/mnras/stu1007}

\bibitem[{{Dijkstra} {et~al.}(2008){Dijkstra}, {Haiman}, {Mesinger}, \&
  {Wyithe}}]{Dijkstra2008}
{Dijkstra}, M., {Haiman}, Z., {Mesinger}, A., \& {Wyithe}, J. S.~B. 2008,
  \mnras, 391, 1961, \dodoi{10.1111/j.1365-2966.2008.14031.x}

\bibitem[{{Eddington}(1916)}]{Edd}
{Eddington}, A.~S. 1916, \mnras, 77, 16, \dodoi{10.1093/mnras/77.1.16}

\bibitem[{{Evans} {et~al.}(2021){Evans}, {Adhikari}, {Afle}, {Ballmer},
  {Biscoveanu}, {Borhanian}, {Brown}, {Chen}, {Eisenstein}, {Gruson}, {Gupta},
  {Hall}, {Huxford}, {Kamai}, {Kashyap}, {Kissel}, {Kuns}, {Landry}, {Lenon},
  {Lovelace}, {McCuller}, {Ng}, {Nitz}, {Read}, {Sathyaprakash}, {Shoemaker},
  {Slagmolen}, {Smith}, {Srivastava}, {Sun}, {Vitale}, \& {Weiss}}]{CE}
{Evans}, M., {Adhikari}, R.~X., {Afle}, C., {et~al.} 2021, arXiv e-prints,
  arXiv:2109.09882, \dodoi{10.48550/arXiv.2109.09882}

\bibitem[{{Ferrara} {et~al.}(2014){Ferrara}, {Salvadori}, {Yue}, \&
  {Schleicher}}]{Ferrara2014}
{Ferrara}, A., {Salvadori}, S., {Yue}, B., \& {Schleicher}, D. 2014, \mnras,
  443, 2410, \dodoi{10.1093/mnras/stu1280}

\bibitem[{{Galli} \& {Palla}(2013)}]{GP2013}
{Galli}, D., \& {Palla}, F. 2013, \araa, 51, 163,
  \dodoi{10.1146/annurev-astro-082812-141029}

\bibitem[{{Greif} {et~al.}(2011){Greif}, {Springel}, {White}, {Glover},
  {Clark}, {Smith}, {Klessen}, \& {Bromm}}]{grief2011}
{Greif}, T.~H., {Springel}, V., {White}, S. D.~M., {et~al.} 2011, \apj, 737,
  75, \dodoi{10.1088/0004-637X/737/2/75}

\bibitem[{{Habouzit} {et~al.}(2016){Habouzit}, {Volonteri}, {Latif}, {Dubois},
  \& {Peirani}}]{Habouzit2016}
{Habouzit}, M., {Volonteri}, M., {Latif}, M., {Dubois}, Y., \& {Peirani}, S.
  2016, \mnras, 463, 529, \dodoi{10.1093/mnras/stw1924}

\bibitem[{{Haiman} \& {Loeb}(2001)}]{haiman2001}
{Haiman}, Z., \& {Loeb}, A. 2001, \apj, 552, 459, \dodoi{10.1086/320586}

\bibitem[{{Haiman} {et~al.}(1997){Haiman}, {Rees}, \& {Loeb}}]{haiman1997}
{Haiman}, Z., {Rees}, M.~J., \& {Loeb}, A. 1997, \apj, 476, 458,
  \dodoi{10.1086/303647}

\bibitem[{{Hills}(1975)}]{hills1975}
{Hills}, J.~G. 1975, \nat, 254, 295, \dodoi{10.1038/254295a0}

\bibitem[{{Hirano} {et~al.}(2015){Hirano}, {Hosokawa}, {Yoshida}, {Omukai}, \&
  {Yorke}}]{Hirano2015}
{Hirano}, S., {Hosokawa}, T., {Yoshida}, N., {Omukai}, K., \& {Yorke}, H.~W.
  2015, \mnras, 448, 568, \dodoi{10.1093/mnras/stv044}

\bibitem[{{Hirano} {et~al.}(2014){Hirano}, {Hosokawa}, {Yoshida}, {Umeda},
  {Omukai}, {Chiaki}, \& {Yorke}}]{Hirano2014}
{Hirano}, S., {Hosokawa}, T., {Yoshida}, N., {et~al.} 2014, \apj, 781, 60,
  \dodoi{10.1088/0004-637X/781/2/60}

\bibitem[{{Hollenbach} \& {McKee}(1979)}]{HM1979}
{Hollenbach}, D., \& {McKee}, C.~F. 1979, \apjs, 41, 555,
  \dodoi{10.1086/190631}

\bibitem[{{Iliev} {et~al.}(2003){Iliev}, {Scannapieco}, {Martel}, \&
  {Shapiro}}]{iliev2003}
{Iliev}, I.~T., {Scannapieco}, E., {Martel}, H., \& {Shapiro}, P.~R. 2003,
  \mnras, 341, 81, \dodoi{10.1046/j.1365-8711.2003.06410.x}

\bibitem[{{Inayoshi} \& {Haiman}(2014)}]{inayoshi2014}
{Inayoshi}, K., \& {Haiman}, Z. 2014, \mnras, 445, 1549,
  \dodoi{10.1093/mnras/stu1870}

\bibitem[{{Inayoshi} {et~al.}(2020){Inayoshi}, {Visbal}, \&
  {Haiman}}]{inayoshi2020}
{Inayoshi}, K., {Visbal}, E., \& {Haiman}, Z. 2020, \araa, 58, 27,
  \dodoi{10.1146/annurev-astro-120419-014455}

\bibitem[{{Jaacks} {et~al.}(2018){Jaacks}, {Thompson}, {Finkelstein}, \&
  {Bromm}}]{Jaacks2018}
{Jaacks}, J., {Thompson}, R., {Finkelstein}, S.~L., \& {Bromm}, V. 2018,
  \mnras, 475, 4396, \dodoi{10.1093/mnras/sty062}

\bibitem[{{Kar Chowdhury} {et~al.}(2024){Kar Chowdhury}, {Chang}, {Dai}, \&
  {Natarajan}}]{Chowdhury2024}
{Kar Chowdhury}, R., {Chang}, J. N.~Y., {Dai}, L., \& {Natarajan}, P. 2024,
  \apjl, 966, L33, \dodoi{10.3847/2041-8213/ad41b7}

\bibitem[{{Karmen} {et~al.}(2025){Karmen}, {Gezari}, {Lambrides}, {Akins},
  {Norman}, {Casey}, {Pierel}, {Coulter}, {Rest}, {Fox}, {Ajay}, {Allen},
  {Drakos}, {Fujimoto}, {Gomez}, {Gozaliasl}, {Ilbert}, {Kartaltepe},
  {Koekemoer}, {Lane}, {McCracken}, {Paquereau}, {Rhodes}, {Robertson},
  {Shuntov}, {Siebert}, {Toft}, {Wevers}, \& {Zenati}}]{karmen2025}
{Karmen}, M., {Gezari}, S., {Lambrides}, E., {et~al.} 2025, arXiv e-prints,
  arXiv:2504.13248.
\newblock \doarXiv{2504.13248}

\bibitem[{{Kashiyama} \& {Inayoshi}(2016)}]{kashiyama2016}
{Kashiyama}, K., \& {Inayoshi}, K. 2016, \apj, 826, 80,
  \dodoi{10.3847/0004-637X/826/1/80}

\bibitem[{{Kiyuna} {et~al.}(2023){Kiyuna}, {Hosokawa}, \& {Chon}}]{K23}
{Kiyuna}, M., {Hosokawa}, T., \& {Chon}, S. 2023, \mnras, 523, 1496,
  \dodoi{10.1093/mnras/stad1484}

\bibitem[{{Kiyuna} {et~al.}(2024){Kiyuna}, {Hosokawa}, \& {Chon}}]{masaki2024}
---. 2024, \mnras, 534, 3916, \dodoi{10.1093/mnras/stae2380}

\bibitem[{{Kocsis} \& {Tremaine}(2011)}]{kocsis2011}
{Kocsis}, B., \& {Tremaine}, S. 2011, \mnras, 412, 187,
  \dodoi{10.1111/j.1365-2966.2010.17897.x}

\bibitem[{{Lacey} \& {Cole}(1993)}]{LC1993}
{Lacey}, C., \& {Cole}, S. 1993, \mnras, 262, 627,
  \dodoi{10.1093/mnras/262.3.627}

\bibitem[{{Lee} {et~al.}(2023){Lee}, {Kim}, \& {Oh}}]{lee2023}
{Lee}, S., {Kim}, J.-h., \& {Oh}, B.~K. 2023, \apj, 943, 77,
  \dodoi{10.3847/1538-4357/aca75d}

\bibitem[{{Li} {et~al.}(2021){Li}, {Inayoshi}, \& {Qiu}}]{Li2021}
{Li}, W., {Inayoshi}, K., \& {Qiu}, Y. 2021, \apj, 917, 60,
  \dodoi{10.3847/1538-4357/ac0adc}

\bibitem[{{Li} {et~al.}(2007){Li}, {Hernquist}, {Robertson}, {Cox}, {Hopkins},
  {Springel}, {Gao}, {Di Matteo}, {Zentner}, {Jenkins}, \& {Yoshida}}]{li2007}
{Li}, Y., {Hernquist}, L., {Robertson}, B., {et~al.} 2007, \apj, 665, 187,
  \dodoi{10.1086/519297}

\bibitem[{{Lodato} \& {Natarajan}(2006)}]{LN2006}
{Lodato}, G., \& {Natarajan}, P. 2006, \mnras, 371, 1813,
  \dodoi{10.1111/j.1365-2966.2006.10801.x}

\bibitem[{{Luo} {et~al.}(2016){Luo}, {Chen}, {Duan}, {Gong}, {Hu}, {Ji}, {Liu},
  {Mei}, {Milyukov}, {Sazhin}, {Shao}, {Toth}, {Tu}, {Wang}, {Wang}, {Yeh},
  {Zhan}, {Zhang}, {Zharov}, \& {Zhou}}]{tianqin}
{Luo}, J., {Chen}, L.-S., {Duan}, H.-Z., {et~al.} 2016, Classical and Quantum
  Gravity, 33, 035010, \dodoi{10.1088/0264-9381/33/3/035010}

\bibitem[{{Lupi} {et~al.}(2014){Lupi}, {Colpi}, {Devecchi}, {Galanti}, \&
  {Volonteri}}]{lupi2014}
{Lupi}, A., {Colpi}, M., {Devecchi}, B., {Galanti}, G., \& {Volonteri}, M.
  2014, \mnras, 442, 3616, \dodoi{10.1093/mnras/stu1120}

\bibitem[{{Lupi} {et~al.}(2016){Lupi}, {Haardt}, {Dotti}, {Fiacconi}, {Mayer},
  \& {Madau}}]{lupi2016}
{Lupi}, A., {Haardt}, F., {Dotti}, M., {et~al.} 2016, \mnras, 456, 2993,
  \dodoi{10.1093/mnras/stv2877}

\bibitem[{{Lupi} {et~al.}(2021){Lupi}, {Haiman}, \& {Volonteri}}]{lupi2021}
{Lupi}, A., {Haiman}, Z., \& {Volonteri}, M. 2021, \mnras, 503, 5046,
  \dodoi{10.1093/mnras/stab692}

\bibitem[{{Ma} {et~al.}(2021){Ma}, {Hopkins}, {Ma}, {Angl{\'e}s-Alc{\'a}zar},
  {Faucher-Gigu{\`e}re}, \& {Kelley}}]{ma2021}
{Ma}, L., {Hopkins}, P.~F., {Ma}, X., {et~al.} 2021, \mnras, 508, 1973,
  \dodoi{10.1093/mnras/stab2713}

\bibitem[{{Madau} \& {Rees}(2001)}]{madau_rees2001}
{Madau}, P., \& {Rees}, M.~J. 2001, \apjl, 551, L27, \dodoi{10.1086/319848}

\bibitem[{{Matsuoka} {et~al.}(2019){Matsuoka}, {Onoue}, {Kashikawa}, {Strauss},
  {Iwasawa}, {Lee}, {Imanishi}, {Nagao}, {Akiyama}, {Asami}, {Bosch},
  {Furusawa}, {Goto}, {Gunn}, {Harikane}, {Ikeda}, {Izumi}, {Kawaguchi},
  {Kato}, {Kikuta}, {Kohno}, {Komiyama}, {Koyama}, {Lupton}, {Minezaki},
  {Miyazaki}, {Murayama}, {Niida}, {Nishizawa}, {Noboriguchi}, {Oguri}, {Ono},
  {Ouchi}, {Price}, {Sameshima}, {Schulze}, {Shirakata}, {Silverman},
  {Sugiyama}, {Tait}, {Takada}, {Takata}, {Tanaka}, {Tang}, {Toba}, {Utsumi},
  {Wang}, \& {Yamashita}}]{matsuoka2019}
{Matsuoka}, Y., {Onoue}, M., {Kashikawa}, N., {et~al.} 2019, \apjl, 872, L2,
  \dodoi{10.3847/2041-8213/ab0216}

\bibitem[{{McCaffrey} {et~al.}(2025){McCaffrey}, {Regan}, {Smith}, {Wise},
  {O'Shea}, \& {Norman}}]{mcCaffrey2025}
{McCaffrey}, J., {Regan}, J., {Smith}, B., {et~al.} 2025, The Open Journal of
  Astrophysics, 8, 11, \dodoi{10.33232/001c.129138}

\bibitem[{{Mehta} {et~al.}(2024){Mehta}, {Regan}, \& {Prole}}]{mehta2024}
{Mehta}, D., {Regan}, J.~A., \& {Prole}, L. 2024, The Open Journal of
  Astrophysics, 7, 107, \dodoi{10.33232/001c.126629}

\bibitem[{{Merritt}(2013)}]{Merritt2013}
{Merritt}, D. 2013, {Dynamics and Evolution of Galactic Nuclei}

\bibitem[{{Milosavljevi{\'c}} {et~al.}(2006){Milosavljevi{\'c}}, {Merritt}, \&
  {Ho}}]{milosavljevic2006}
{Milosavljevi{\'c}}, M., {Merritt}, D., \& {Ho}, L.~C. 2006, \apj, 652, 120,
  \dodoi{10.1086/508134}

\bibitem[{{Mo} {et~al.}(1996){Mo}, {Jing}, \& {White}}]{Mo1996}
{Mo}, H.~J., {Jing}, Y.~P., \& {White}, S.~D.~M. 1996, \mnras, 282, 1096,
  \dodoi{10.1093/mnras/282.3.1096}

\bibitem[{{Montero} {et~al.}(2012){Montero}, {Janka}, \&
  {M{\"u}ller}}]{montero2012}
{Montero}, P.~J., {Janka}, H.-T., \& {M{\"u}ller}, E. 2012, \apj, 749, 37,
  \dodoi{10.1088/0004-637X/749/1/37}

\bibitem[{{Mortlock} {et~al.}(2011){Mortlock}, {Warren}, {Venemans}, {Patel},
  {Hewett}, {McMahon}, {Simpson}, {Theuns}, {Gonz{\'a}les-Solares}, {Adamson},
  {Dye}, {Hambly}, {Hirst}, {Irwin}, {Kuiper}, {Lawrence}, \&
  {R{\"o}ttgering}}]{mortlock2011}
{Mortlock}, D.~J., {Warren}, S.~J., {Venemans}, B.~P., {et~al.} 2011, \nat,
  474, 616, \dodoi{10.1038/nature10159}

\bibitem[{{Natarajan} {et~al.}(2017){Natarajan}, {Pacucci}, {Ferrara},
  {Agarwal}, {Ricarte}, {Zackrisson}, \& {Cappelluti}}]{natarajan2017}
{Natarajan}, P., {Pacucci}, F., {Ferrara}, A., {et~al.} 2017, \apj, 838, 117,
  \dodoi{10.3847/1538-4357/aa6330}

\bibitem[{{Oh} \& {Haiman}(2002)}]{Oh2002}
{Oh}, S.~P., \& {Haiman}, Z. 2002, \apj, 569, 558, \dodoi{10.1086/339393}

\bibitem[{{Omukai} \& {Palla}(2003)}]{OP2003}
{Omukai}, K., \& {Palla}, F. 2003, \apj, 589, 677, \dodoi{10.1086/374810}

\bibitem[{{Pacucci} {et~al.}(2015){Pacucci}, {Ferrara}, {Volonteri}, \&
  {Dubus}}]{pacucci2015}
{Pacucci}, F., {Ferrara}, A., {Volonteri}, M., \& {Dubus}, G. 2015, \mnras,
  454, 3771, \dodoi{10.1093/mnras/stv2196}

\bibitem[{{Parkinson} {et~al.}(2008){Parkinson}, {Cole}, \&
  {Helly}}]{parkinson2008}
{Parkinson}, H., {Cole}, S., \& {Helly}, J. 2008, \mnras, 383, 557,
  \dodoi{10.1111/j.1365-2966.2007.12517.x}

\bibitem[{{Pfister} {et~al.}(2021){Pfister}, {Dai}, {Volonteri}, {Auchettl},
  {Trebitsch}, \& {Ramirez-Ruiz}}]{pfister2021}
{Pfister}, H., {Dai}, J.~L., {Volonteri}, M., {et~al.} 2021, \mnras, 500, 3944,
  \dodoi{10.1093/mnras/staa3471}

\bibitem[{{Pfister} {et~al.}(2019){Pfister}, {Volonteri}, {Dubois}, {Dotti}, \&
  {Colpi}}]{pfister2019}
{Pfister}, H., {Volonteri}, M., {Dubois}, Y., {Dotti}, M., \& {Colpi}, M. 2019,
  \mnras, 486, 101, \dodoi{10.1093/mnras/stz822}

\bibitem[{{Planck Collaboration} {et~al.}(2020){Planck Collaboration},
  {Aghanim}, {Akrami}, {Ashdown}, {Aumont}, {Baccigalupi}, {Ballardini},
  {Banday}, {Barreiro}, {Bartolo}, {Basak}, {Battye}, {Benabed}, {Bernard},
  {Bersanelli}, {Bielewicz}, {Bock}, {Bond}, {Borrill}, {Bouchet}, {Boulanger},
  {Bucher}, {Burigana}, {Butler}, {Calabrese}, {Cardoso}, {Carron},
  {Challinor}, {Chiang}, {Chluba}, {Colombo}, {Combet}, {Contreras}, {Crill},
  {Cuttaia}, {de Bernardis}, {de Zotti}, {Delabrouille}, {Delouis}, {Di
  Valentino}, {Diego}, {Dor{\'e}}, {Douspis}, {Ducout}, {Dupac}, {Dusini},
  {Efstathiou}, {Elsner}, {En{\ss}lin}, {Eriksen}, {Fantaye}, {Farhang},
  {Fergusson}, {Fernandez-Cobos}, {Finelli}, {Forastieri}, {Frailis},
  {Fraisse}, {Franceschi}, {Frolov}, {Galeotta}, {Galli}, {Ganga},
  {G{\'e}nova-Santos}, {Gerbino}, {Ghosh}, {Gonz{\'a}lez-Nuevo}, {G{\'o}rski},
  {Gratton}, {Gruppuso}, {Gudmundsson}, {Hamann}, {Handley}, {Hansen},
  {Herranz}, {Hildebrandt}, {Hivon}, {Huang}, {Jaffe}, {Jones}, {Karakci},
  {Keih{\"a}nen}, {Keskitalo}, {Kiiveri}, {Kim}, {Kisner}, {Knox},
  {Krachmalnicoff}, {Kunz}, {Kurki-Suonio}, {Lagache}, {Lamarre}, {Lasenby},
  {Lattanzi}, {Lawrence}, {Le Jeune}, {Lemos}, {Lesgourgues}, {Levrier},
  {Lewis}, {Liguori}, {Lilje}, {Lilley}, {Lindholm}, {L{\'o}pez-Caniego},
  {Lubin}, {Ma}, {Mac{\'\i}as-P{\'e}rez}, {Maggio}, {Maino}, {Mandolesi},
  {Mangilli}, {Marcos-Caballero}, {Maris}, {Martin}, {Martinelli},
  {Mart{\'\i}nez-Gonz{\'a}lez}, {Matarrese}, {Mauri}, {McEwen}, {Meinhold},
  {Melchiorri}, {Mennella}, {Migliaccio}, {Millea}, {Mitra},
  {Miville-Desch{\^e}nes}, {Molinari}, {Montier}, {Morgante}, {Moss}, {Natoli},
  {N{\o}rgaard-Nielsen}, {Pagano}, {Paoletti}, {Partridge}, {Patanchon},
  {Peiris}, {Perrotta}, {Pettorino}, {Piacentini}, {Polastri}, {Polenta},
  {Puget}, {Rachen}, {Reinecke}, {Remazeilles}, {Renzi}, {Rocha}, {Rosset},
  {Roudier}, {Rubi{\~n}o-Mart{\'\i}n}, {Ruiz-Granados}, {Salvati}, {Sandri},
  {Savelainen}, {Scott}, {Shellard}, {Sirignano}, {Sirri}, {Spencer},
  {Sunyaev}, {Suur-Uski}, {Tauber}, {Tavagnacco}, {Tenti}, {Toffolatti},
  {Tomasi}, {Trombetti}, {Valenziano}, {Valiviita}, {Van Tent}, {Vibert},
  {Vielva}, {Villa}, {Vittorio}, {Wandelt}, {Wehus}, {White}, {White},
  {Zacchei}, \& {Zonca}}]{planck2020}
{Planck Collaboration}, {Aghanim}, N., {Akrami}, Y., {et~al.} 2020, \aap, 641,
  A6, \dodoi{10.1051/0004-6361/201833910}

\bibitem[{{Press} \& {Schechter}(1974)}]{PS}
{Press}, W.~H., \& {Schechter}, P. 1974, \apj, 187, 425, \dodoi{10.1086/152650}

\bibitem[{{Regan} {et~al.}(2014){Regan}, {Johansson}, \&
  {Haehnelt}}]{regan2014}
{Regan}, J.~A., {Johansson}, P.~H., \& {Haehnelt}, M.~G. 2014, \mnras, 439,
  1160, \dodoi{10.1093/mnras/stu068}

\bibitem[{{Ricarte} \& {Natarajan}(2018)}]{RN18a}
{Ricarte}, A., \& {Natarajan}, P. 2018, \mnras, 481, 3278,
  \dodoi{10.1093/mnras/sty2448}

\bibitem[{Rigby {et~al.}(2023)Rigby, Perrin, McElwain, Kimble, Friedman, Lallo,
  Doyon, Feinberg, Ferruit, Glasse, Rieke, Rieke, Wright, Willott, Colon,
  Milam, Neff, Stark, Valenti, Abell, Abney, Abul-Huda, Scott~Acton, Adams,
  Adler, Aguilar, Ahmed, Albert, Alberts, Aldridge, Allen, Altenburg, Álvarez
  Márquez, Alves~de Oliveira, Andersen, Anderson, Anderson, Argyriou,
  Armstrong, Arribas, Artigau, Arvai, Atkinson, Bacon, Bair, Banks, Barrientes,
  Barringer, Bartosik, Bast, Baudoz, Beatty, Bechtold, Beck, Bergeron,
  Bergkoetter, Bhatawdekar, Birkmann, Blazek, Blome, Boccaletti, Böker, Boia,
  Bonaventura, Bond, Bosley, Boucarut, Bourque, Bouwman, Bower, Bowers, Boyer,
  Bradley, Brady, Braun, Breda, Bresnahan, Bright, Britt, Bromenschenkel,
  Brooks, Brooks, Brown, Brown, Brown, Bunker, Burger, Bushouse, Cale, Cameron,
  Cameron, Canipe, Caplinger, Caputo, Cara, Carey, Carniani, Carrasquilla,
  Carruthers, Case, Catherine, Chance, Chapman, Charlot, Charlow, Chayer, Chen,
  Cherinka, Chichester, Chilton, Chonis, Clampin, Clark, Clark, Coe, Coleman,
  Comber, Comeau, Connolly, Cooper, Cooper, Coppock, Correnti, Cossou, Coulais,
  Coyle, Cracraft, Curti, Cuturic, Davis, Davis, Dean, DeLisa, deMeester,
  Dencheva, Dencheva, DePasquale, Deschenes, Hunor~Detre, Diaz, Dicken,
  DiFelice, Dillman, Dixon, Doggett, Donaldson, Douglas, DuPrie, Dupuis,
  Durning, Easmin, Eck, Edeani, Egami, Ehrenwinkler, Eisenhamer, Eisenhower,
  Elie, Elliott, Elliott, Ellis, Engesser, Espinoza, Etienne, Etxaluze, Falini,
  Feeney, Ferry, Filippazzo, Fincham, Fix, Flagey, Florian, Flynn, Fontanella,
  Ford, Forshay, Fox, Franz, Fu, Fullerton, Galkin, Galyer, García~Marín,
  Gardner, Gardner, Garland, Garrett, Gasman, Gaspar, Gaudreau, Gauthier,
  Geers, Geithner, Gennaro, Giardino, Girard, Giuliano, Glassmire, Glauser,
  Glazer, Godfrey, Golimowski, Gollnitz, Gong, Gonzaga, Gordon, Gordon,
  Goudfrooij, Greene, Greenhouse, Grimaldi, Groebner, Grundy, Guillard, Gutman,
  Ha, Haderlein, Hagedorn, Hainline, Haley, Hami, Hamilton, Hammel, Hansen,
  Harkins, Harr, Hart, Hart, Hartig, Hashimoto, Haskins, Hathaway, Havey,
  Hayden, Hecht, Heller-Boyer, Henriques, Henry, Hermann, Hernandez, Hesman,
  Hicks, Hilbert, Hines, Hoffman, Holfeltz, Holler, Hoppa, Hott, Howard,
  Howard, Hunter, Hunter, Hurst, Husemann, Hustak, Ilinca~Ignat, Illingworth,
  Irish, Jackson, Jahromi, Jakobsen, James, James, Januszewski, Jenkins,
  Jirdeh, Johnson, Johnson, Jones, Jones, Jones, Jones, Jordan, Jordan,
  Jurczyk, Jurling, Kaleida, Kalmanson, Kammerer, Kang, Kao, Karakla, Kavanagh,
  Kelly, Kendrew, Kennedy, Kenny, Keski-kuha, Keyes, Kidwell, Kinzel, Kirk,
  Kirkpatrick, Kirshenblat, Klaassen, Knapp, Scott~Knight, Knollenberg,
  Koehler, Koekemoer, Kovacs, Kulp, Kumari, Kyprianou, La~Massa, Labador,
  Labiano, Lagage, Lajoie, Lallo, Lam, Lamb, Lambros, Lampenfield, Langston,
  Larson, Law, Lawrence, Lee, Leisenring, Lepo, Leveille, Levenson, Levine,
  Levy, Lewis, Lewis, Libralato, Lightsey, Link, Liu, Lo, Lockwood, Logue,
  Long, Long, Loomis, Lopez-Caniego, Lorenzo~Alvarez, Love-Pruitt, Lucy,
  Luetzgendorf, Maghami, Maiolino, Major, Malla, Malumuth, Manjavacas,
  Mannfolk, Marrione, Marston, Martel, Maschmann, Masci, Masciarelli,
  Maszkiewicz, Mather, McKenzie, McLean, McMaster, Melbourne, Meléndez,
  Menzel, Merz, Meyett, Meza, Miskey, Misselt, Moller, Morrison, Morse,
  Moseley, Mosier, Mountain, Mueckay, Mueller, Mullally, Murphy, Murray,
  Murray, Mustelier, Muzerolle, Mycroft, Myers, Myrick, Nanavati, Nance, Nayak,
  Naylor, Nelan, Nickson, Nielson, Nieto-Santisteban, Nikolov, Noriega-Crespo,
  O’Shaughnessy, O’Sullivan, Ochs, Ogle, Oleszczuk, Olmsted, Osborne,
  Ottens, Owens, Pacifici, Pagan, Page, Park, Parrish, Patapis, Paul, Pauly,
  Pavlovsky, Pedder, Peek, Pena-Guerrero, Penanen, Perez, Perna, Perriello,
  Phillips, Pietraszkiewicz, Pinaud, Pirzkal, Pitman, Piwowar, Platais, Player,
  Plesha, Pollizi, Polster, Pontoppidan, Porterfield, Proffitt, Pueyo, Pulliam,
  Quirt, Quispe~Neira, Ramos~Alarcon, Ramsay, Rapp, Rapp, Rauscher,
  Ravindranath, Rawle, Regan, Reichard, Reis, Ressler, Rest, Reynolds, Rhue,
  Richon, Rickman, Ridgaway, Ritchie, Rix, Robberto, Robinson, Robinson,
  Robinson, Rock, Rodriguez, Rodriguez Del~Pino, Roellig, Rohrbach, Roman,
  Romelfanger, Rose, Roteliuk, Roth, Rothwell, Rowlands, Roy, Royer, Royle,
  Rui, Rumler, Runnels, Russ, Rustamkulov, Ryden, Ryer, Sabata, Sabatke, Sabbi,
  Samuelson, Sapp, Sappington, Sargent, Sauer, Scheithauer, Schlawin, Schlitz,
  Schmitz, Schneider, Schreiber, Schulze, Schwab, Scott, Sembach, Shanahan,
  Shaughnessy, Shaw, Shawger, Shay, Sheehan, Shen, Sherman, Shiao, Shih,
  Shivaei, Sienkiewicz, Sing, Sirianni, Sivaramakrishnan, Skipper, Sloan,
  Slocum, Slowinski, Smith, Smith, Smith, Smith, Snyder, Soh, Tony~Sohn, Soto,
  Spencer, Stallcup, Stansberry, Starr, Starr, Stewart, Stiavelli, Straughn,
  Strickland, Stys, Summers, Sun, Sunnquist, Swade, Swam, Swaters, Swoish,
  Taylor, Taylor, Te~Plate, Tea, Teague, Telfer, Temim, Thatte, Thompson,
  Thompson, Thomson, Tikkanen, Tippet, Todd, Toolan, Tran, Trejo, Truong,
  Tsukamoto, Tustain, Tyra, Ubeda, Underwood, Uzzo, Van~Campen, Vandal,
  Vandenbussche, Vila, Volk, Wahlgren, Waldman, Walker, Wander, Warfield,
  Warner, Wasiak, Watkins, Weaver, Weilert, Weiser, Weiss, Weissman, Welty,
  West, Wheate, Wheatley, Wheeler, White, Whiteaker, Whitehouse, Whiteleather,
  Whitman, Williams, Willmer, Willoughby, Wilson, Wirth, Wislowski, Wolf,
  Wolfe, Wolff, Workman, Wright, Wu, Wu, Wymer, Yates, Yeager, Yeates, Yerger,
  Yoon, Young, Yu, Zak, Zeidler, Zhou, Zielinski, Zincke, \&
  Zonak}]{Rigby_2023}
Rigby, J., Perrin, M., McElwain, M., {et~al.} 2023, Publications of the
  Astronomical Society of the Pacific, 135, 048001,
  \dodoi{10.1088/1538-3873/acb293}

\bibitem[{{Rizzuto} {et~al.}(2023){Rizzuto}, {Naab}, {Rantala}, {Johansson},
  {Ostriker}, {Stone}, {Liao}, \& {Irodotou}}]{Rizzuto2023}
{Rizzuto}, F.~P., {Naab}, T., {Rantala}, A., {et~al.} 2023, \mnras, 521, 2930,
  \dodoi{10.1093/mnras/stad734}

\bibitem[{{Salpeter}(1955)}]{salpeter1955}
{Salpeter}, E.~E. 1955, \apj, 121, 161, \dodoi{10.1086/145971}

\bibitem[{{Scannapieco} \& {Barkana}(2002)}]{SB2002}
{Scannapieco}, E., \& {Barkana}, R. 2002, \apj, 571, 585,
  \dodoi{10.1086/340063}

\bibitem[{{Schaerer}(2002)}]{schaerer2002}
{Schaerer}, D. 2002, \aap, 382, 28, \dodoi{10.1051/0004-6361:20011619}

\bibitem[{{Schneider} {et~al.}(2002){Schneider}, {Ferrara}, {Natarajan}, \&
  {Omukai}}]{schneider2002}
{Schneider}, R., {Ferrara}, A., {Natarajan}, P., \& {Omukai}, K. 2002, \apj,
  571, 30, \dodoi{10.1086/339917}

\bibitem[{{Scoggins} \& {Haiman}(2024)}]{scoggins2024}
{Scoggins}, M.~T., \& {Haiman}, Z. 2024, \mnras, 531, 4584,
  \dodoi{10.1093/mnras/stae1449}

\bibitem[{{Shang} {et~al.}(2010){Shang}, {Bryan}, \& {Haiman}}]{shang2010}
{Shang}, C., {Bryan}, G.~L., \& {Haiman}, Z. 2010, \mnras, 402, 1249,
  \dodoi{10.1111/j.1365-2966.2009.15960.x}

\bibitem[{{Shen} {et~al.}(2023){Shen}, {Vogelsberger}, {Boylan-Kolchin},
  {Tacchella}, \& {Kannan}}]{shen2023}
{Shen}, X., {Vogelsberger}, M., {Boylan-Kolchin}, M., {Tacchella}, S., \&
  {Kannan}, R. 2023, \mnras, 525, 3254, \dodoi{10.1093/mnras/stad2508}

\bibitem[{{Shi} {et~al.}(2023){Shi}, {Kremer}, {Grudi{\'c}},
  {Gerling-Dunsmore}, \& {Hopkins}}]{shi2023}
{Shi}, Y., {Kremer}, K., {Grudi{\'c}}, M.~Y., {Gerling-Dunsmore}, H.~J., \&
  {Hopkins}, P.~F. 2023, \mnras, 518, 3606, \dodoi{10.1093/mnras/stac3245}

\bibitem[{{Spergel} {et~al.}(2015){Spergel}, {Gehrels}, {Baltay}, {Bennett},
  {Breckinridge}, {Donahue}, {Dressler}, {Gaudi}, {Greene}, {Guyon}, {Hirata},
  {Kalirai}, {Kasdin}, {Macintosh}, {Moos}, {Perlmutter}, {Postman},
  {Rauscher}, {Rhodes}, {Wang}, {Weinberg}, {Benford}, {Hudson}, {Jeong},
  {Mellier}, {Traub}, {Yamada}, {Capak}, {Colbert}, {Masters}, {Penny},
  {Savransky}, {Stern}, {Zimmerman}, {Barry}, {Bartusek}, {Carpenter}, {Cheng},
  {Content}, {Dekens}, {Demers}, {Grady}, {Jackson}, {Kuan}, {Kruk}, {Melton},
  {Nemati}, {Parvin}, {Poberezhskiy}, {Peddie}, {Ruffa}, {Wallace}, {Whipple},
  {Wollack}, \& {Zhao}}]{spergel2015}
{Spergel}, D., {Gehrels}, N., {Baltay}, C., {et~al.} 2015, arXiv e-prints,
  arXiv:1503.03757, \dodoi{10.48550/arXiv.1503.03757}

\bibitem[{{Spitzer}(1987)}]{spitzer1987}
{Spitzer}, L. 1987, {Dynamical evolution of globular clusters}

\bibitem[{{Stacy} \& {Bromm}(2013)}]{stacy2013}
{Stacy}, A., \& {Bromm}, V. 2013, \mnras, 433, 1094,
  \dodoi{10.1093/mnras/stt789}

\bibitem[{{Stewart} \& {Ida}(2000)}]{stewart2000}
{Stewart}, G.~R., \& {Ida}, S. 2000, \icarus, 143, 28,
  \dodoi{10.1006/icar.1999.6242}

\bibitem[{{Tanaka} {et~al.}(2013){Tanaka}, {Nakamoto}, \&
  {Omukai}}]{tanaka2013}
{Tanaka}, K. E.~I., {Nakamoto}, T., \& {Omukai}, K. 2013, \apj, 773, 155,
  \dodoi{10.1088/0004-637X/773/2/155}

\bibitem[{{Trinca} {et~al.}(2023){Trinca}, {Schneider}, {Maiolino}, {Valiante},
  {Graziani}, \& {Volonteri}}]{trinca2023}
{Trinca}, A., {Schneider}, R., {Maiolino}, R., {et~al.} 2023, \mnras, 519,
  4753, \dodoi{10.1093/mnras/stac3768}

\bibitem[{{Valiante} {et~al.}(2018){Valiante}, {Schneider}, {Zappacosta},
  {Graziani}, {Pezzulli}, \& {Volonteri}}]{Valiante2018}
{Valiante}, R., {Schneider}, R., {Zappacosta}, L., {et~al.} 2018, \mnras, 476,
  407, \dodoi{10.1093/mnras/sty213}

\bibitem[{{Volonteri} {et~al.}(2003){Volonteri}, {Haardt}, \&
  {Madau}}]{volonteri2003}
{Volonteri}, M., {Haardt}, F., \& {Madau}, P. 2003, \apj, 582, 559,
  \dodoi{10.1086/344675}

\bibitem[{{Volonteri} {et~al.}(2021){Volonteri}, {Habouzit}, \&
  {Colpi}}]{volonteri2021}
{Volonteri}, M., {Habouzit}, M., \& {Colpi}, M. 2021, Nature Reviews Physics,
  3, 732, \dodoi{10.1038/s42254-021-00364-9}

\bibitem[{{Wang} {et~al.}(2018){Wang}, {Yang}, {Fan}, {Yue}, {Wu}, {Schindler},
  {Bian}, {Li}, {Farina}, {Ba{\~n}ados}, {Davies}, {Decarli}, {Green}, {Jiang},
  {Hennawi}, {Huang}, {Mazzucchelli}, {McGreer}, {Venemans}, {Walter}, \&
  {Beletsky}}]{wang2018}
{Wang}, F., {Yang}, J., {Fan}, X., {et~al.} 2018, \apjl, 869, L9,
  \dodoi{10.3847/2041-8213/aaf1d2}

\bibitem[{{Wang} {et~al.}(2020){Wang}, {Davies}, {Yang}, {Hennawi}, {Fan},
  {Barth}, {Jiang}, {Wu}, {Mudd}, {Ba{\~n}ados}, {Bian}, {Decarli}, {Eilers},
  {Farina}, {Venemans}, {Walter}, \& {Yue}}]{wang2020}
{Wang}, F., {Davies}, F.~B., {Yang}, J., {et~al.} 2020, \apj, 896, 23,
  \dodoi{10.3847/1538-4357/ab8c45}

\bibitem[{{Wang} {et~al.}(2021){Wang}, {Yang}, {Fan}, {Hennawi}, {Barth},
  {Banados}, {Bian}, {Boutsia}, {Connor}, {Davies}, {Decarli}, {Eilers},
  {Farina}, {Green}, {Jiang}, {Li}, {Mazzucchelli}, {Nanni}, {Schindler},
  {Venemans}, {Walter}, {Wu}, \& {Yue}}]{wang2021}
{Wang}, F., {Yang}, J., {Fan}, X., {et~al.} 2021, \apjl, 907, L1,
  \dodoi{10.3847/2041-8213/abd8c6}

\bibitem[{Wang {et~al.}(2020)Wang, Ni, Han, Yang, \& Zhong}]{taiji}
Wang, G., Ni, W.-T., Han, W.-B., Yang, S.-C., \& Zhong, X.-Y. 2020, Phys. Rev.
  D, 102, 024089, \dodoi{10.1103/PhysRevD.102.024089}

\bibitem[{{Whalen} {et~al.}(2020){Whalen}, {Surace}, {Bernhardt}, {Zackrisson},
  {Pacucci}, {Ziegler}, \& {Hirschmann}}]{whalen2020}
{Whalen}, D.~J., {Surace}, M., {Bernhardt}, C., {et~al.} 2020, \apjl, 897, L16,
  \dodoi{10.3847/2041-8213/ab9d29}

\bibitem[{{Wise} {et~al.}(2012){Wise}, {Turk}, {Norman}, \& {Abel}}]{wise2012}
{Wise}, J.~H., {Turk}, M.~J., {Norman}, M.~L., \& {Abel}, T. 2012, \apj, 745,
  50, \dodoi{10.1088/0004-637X/745/1/50}

\bibitem[{{Wolcott-Green} {et~al.}(2017){Wolcott-Green}, {Haiman}, \&
  {Bryan}}]{Wolcott-Green2017}
{Wolcott-Green}, J., {Haiman}, Z., \& {Bryan}, G.~L. 2017, \mnras, 469, 3329,
  \dodoi{10.1093/mnras/stx167}

\bibitem[{{Wyithe} \& {Padmanabhan}(2006)}]{wyithe2006}
{Wyithe}, J. S.~B., \& {Padmanabhan}, T. 2006, \mnras, 366, 1029,
  \dodoi{10.1111/j.1365-2966.2005.09858.x}

\bibitem[{{Yang} {et~al.}(2020){Yang}, {Wang}, {Fan}, {Hennawi}, {Davies},
  {Yue}, {Banados}, {Wu}, {Venemans}, {Barth}, {Bian}, {Boutsia}, {Decarli},
  {Farina}, {Green}, {Jiang}, {Li}, {Mazzucchelli}, \& {Walter}}]{yang2020}
{Yang}, J., {Wang}, F., {Fan}, X., {et~al.} 2020, \apjl, 897, L14,
  \dodoi{10.3847/2041-8213/ab9c26}

\bibitem[{{Yoshida} {et~al.}(2003){Yoshida}, {Abel}, {Hernquist}, \&
  {Sugiyama}}]{yoshida2003}
{Yoshida}, N., {Abel}, T., {Hernquist}, L., \& {Sugiyama}, N. 2003, \apj, 592,
  645, \dodoi{10.1086/375810}

\end{thebibliography}
\bibliographystyle{aasjournal}



\end{document}